\documentclass[aps,pre,twocolumn,superscriptaddress]{revtex4-1}
\usepackage{graphicx}
\usepackage{xcolor}
\usepackage{dcolumn}
\usepackage{bm}
\usepackage{amsmath}
\usepackage{amssymb}

\begin{document}

%Title of paper
\title{Growth and shrinkage of tissue sheets on substrates: buds, buckles, and pores}

\author{Hiroshi Noguchi}
\email[]{noguchi@issp.u-tokyo.ac.jp}
\affiliation{Institute for Solid State Physics, University of Tokyo, Kashiwa, Chiba 277-8581, Japan}

\author{Jens Elgeti}
\affiliation{Theoretical Physics of Living Matter, Institute of Biological Information Processing and Institute for Advanced Simulation, Forschungszentrum J\"ulich, 52425 J\"ulich, Germany}

%\date{\today}

\begin{abstract}
Many tissues take the form of thin sheets, being only a single cell thick, but millions of cells wide. 
These tissue sheets can bend and buckle in the third dimension. 
In this work, we investigated the growth and shrinkage of suspended and supported tissue sheets using particle-based simulations. We construct a minimum model, combining particle-based tissue growth and meshless membrane models, to simulate the growth of tissue sheets with mechanical feedback.
Free suspended growing tissues exhibit wrinkling when growth is sufficiently fast.
Conversely, tissues on a substrate form buds when the adhesion to the substrate is weak and/or when the friction with the substrate is strong. These buds undergo a membrane-mediated attraction and subsequently fuse. The complete detachment of tissues from the substrate and straight buckled bump formation are also obtained at very weak adhesion and/or fast growth rates. In the tissue shrinkage, tissue pores grow via Ostwald ripening and coalescence. The reported dynamics can also be applied in research on the detachment dynamics of different tissues with weakened adhesion.
\end{abstract}

\maketitle

\section{Introduction}

Epithelia are thin tissues that serve as barriers around organs and to the outside world. While the most famous epithelium, the skin,  is several cell layers thick, most epithelia are actually single-layered or just a few cells thick. All epithelia are much more extended in the lateral than in the perpendicular directions, motivating a two-dimensional (2D) model. 
However, a 2D model hides important tissue deformations in the third dimension. 
For example, the intestinal epithelium that lines the gut exhibits characteristic finger-like protrusions (called villi) and invaginations (called crypts)~\cite{beum21,shye13}.
A second prominent example of a quasi-two-dimensional tissue growth leading to a three-dimensional (3D) structure occurs in the brain cortex: the differential growth hypothesis states that it is the lateral growth of the gray matter that leads to buckling and thus to the folding of the brain~\cite{tall16}.
The theoretical modeling of quasi-2D sheets growing and embedded in 3D space has proven to be challenging. 
Analytical modeling is usually limited to small deformations, thus one has to revert to numerical simulations.

Simulations of continuum models (e.g., finite element method) are useful for investigating the mechanical responses of tissues.
However, the discrete nature of cells has important consequences, particularly in the fluctuations, evolution, or the probability of jackpot events \cite{hallatschekProliferatingActiveMatter2023}, since cell division plays an essential role.
Therefore, three types of cell-based models have been developed for tissue mechanics involving cell deformation and proliferation:
triangulated cell models~\cite{lied20,okud23,cuve23,ichb23,runs24}, cell vertex models~\cite{flet14,alt17}, and two-particle cell models~\cite{basan2011}.
Triangulated cell models can describe detailed cell shapes and internal structures but are not suitable for large-scale simulations due to their high numerical costs.
Cell vertex models describe a cell as a polygon~\cite{muri15,luci21,ogit22,tetl19,prak21,xu22,sona23,lv24,guer23,hira24} or polyhedron~\cite{hond04,okud13,kraj18,okud18,inou20,zhan22} and
widely used for large-scale tissue growth and deformation. Overdamped Langevin equations are typically used for cell motions so that the hydrodynamic interactions are neglected.
In cell-centered models~\cite{mein01,dape16,mimu23} (variants of the cell vertex models), the cell-center positions are used as the degrees of freedom, and the cell vertices are generated by the Voronoi tessellation; the same potential energy is used as in the vertex model.

In this study, we use the third type of cell models, the two-particle cell models~\cite{basan2011}.
This type of models were developed to explore the role of mechanical feedback on growing matter. The basic idea is that cells can grow (i.e., expand in volume) under a finite force and fluctuation, and the cells divide when a critical size is reached \cite{basan2011}. 
Owing to their simplicity, these models have been applied to various phenomena, such as fluidization of tissues~\cite{ranftFluidizationTissuesCell2010}, growth response of cancer spheroids to pressure \cite{montelStressClampExperiments2011, montelIsotropicStressReduces2012, delarueMechanicalControlCell2013}, and competition and evolution of tissues \cite{buscherTissueEvolutionMechanical2020, buscherInstabilityFingeringInterfaces2020, ganaiMechanicsTissueCompetition2019,podewitzInterfaceDynamicsCompeting2016}. 
Thanks to the particle-based nature, it was easily extended to include cellular motility \cite{basanAlignmentCellularMotility2013, marelAlignmentCellDivision2014}, stem cell dynamics \cite{kramerMechanicallydrivenStemCell2024} or even match bacterial growth \cite{hornungQuantitativeModellingNutrientlimited2018}.

In this work, we combine the two-particle growth model with a particle-based meshless membrane model~\cite{nogu06} to create a minimum model to simulate tissue sheets.
The meshless membrane models were originally developed to simulate membrane dynamics accompanied by topological changes~\cite{nogu10,drou91,nogu06}, and were applied to the vesicle formation~\cite{drou91,nogu06a}, the membrane tubulation induced by curvature-inducing proteins~\cite{nogu22a,nogu16,nogu22b}, the membrane detachment from a substrate~\cite{nogu19c}, and the assembly by inter-membrane entropic repulsion~\cite{nogu13}.

\begin{figure*}
\includegraphics[]{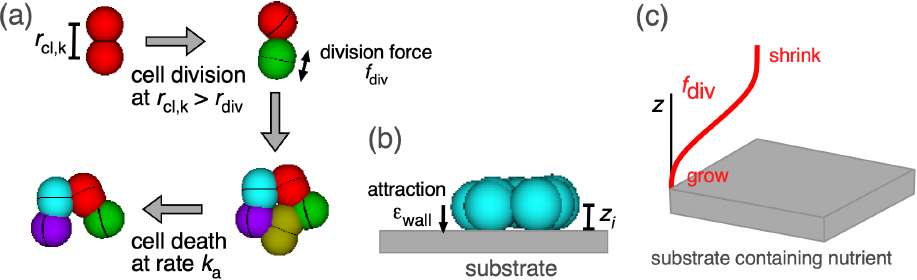}
\caption{
Schematic of the simulation model and setup.
(a) A cell consists of two particles, with a repulsive force $f_{\mathrm{div}}$ working between them.
The cell divides into two cells at $r_{\mathrm{cl},k}>r_{\mathrm{div}}$
 and stochastically dies at the rate $k_{\mathrm{a}}$.
Different cells are displayed in different colors.
(b) A tissue adheres to a substrate wall through an attraction potential with the strength $\varepsilon_{\mathrm{wall}}$.
(c) When simulating the tissues shrinking far from a substrate (Sec.~\ref{sec:steady}),
$f_{\mathrm{div}}$ is set to be a function of the vertical position $z$.
}
\label{fig:cart}
\end{figure*}

The aim of this study is  to simulate the tissue dynamics on an adhesive substrate
and characterize the growth and shrinkage of quai-2D tissue sheets in 3D space. 
We use a solid flat substrate to model a gel sheet in cultured-tissue experiments or hard tissue in a living body.
In the skin, blisters are formed by various diseases (e.g., varicella~\cite{gers15}) and wounding (e.g., burns and rubbing~\cite{hsu18}). 
We investigate the essential detachment mechanism using simple geometry and conditions. We focus our analyses on how the stress in tissue proliferation induces local or entire detachment and how detachment develops.
Previously, Okuda et al.~\cite{okud18} simulated the formation of multiple cylindrical buds from growing spots in a spherical tissue; however, they did not observe bud fusion. Here, we demonstrate that buds fuse through attractive interaction generated by tissue bending.

The simulation model and method are described in Sec.~\ref{sec:method}.
The dynamics of a freely suspended tissue sheets are presented in Sec.~\ref{sec:free}.
The dynamics of growing tissues on a substrate with constant and height-dependent division
are presented and discussed in Secs.~\ref{sec:disk} and \ref{sec:steady}, respectively.
The pore formation in flat tissues is investigated in Sec.~\ref{sec:pore}.
Finally, the conclusions are presented in Sec.~\ref{sec:sum}.

\section{Model and method}\label{sec:method}

The cell growth mechanism is the same as that of the original 3D model~\cite{basan2011}.
Each cell consists of two particles and grows by a repulsive force of magnitude $f_{\mathrm {div}}$ between the two particles; for the $k$-th cell, the two particles push each other as 
$\pm f_{\mathrm {div}} \hat{\mathbf{r}}_{\mathrm{cl},k}$,
where $\hat{\mathbf{r}}_{\mathrm{cl},k}=  \mathbf{r}_{\mathrm{cl},k}/r_{\mathrm{cl},k}$,
 $r_{\mathrm{cl,k}}=|\mathbf{r}_{\mathrm{cl},k}|$,  $\mathbf{r}_{\mathrm{cl},k} = \mathbf{r}_i-\mathbf{r}_j$, and 
$\mathbf{r}_i$ and $\mathbf{r}_j$  are the positions of the two particles in the cell. 
When the distance $r_{\mathrm{cl},k}$ exceeds the threshold value $r_{\mathrm{div}}$,
the $k$-th cell exhibits division into two daughter cells; the centers of two daughter cells locale at the original positions of the particles of the mother cell. 
The two particles in each daughter cell are placed $0.001r_{\mathrm{div}}$ in a random direction from the cell center [see Fig.~\ref{fig:cart}(a)].
The two particles of each daughter cell have the same velocity as the divided particle in the mother cell.
Cells are eliminated at the rate $k_{\mathrm{a}}$ as cell death by either apoptosis or necrosis.
We set the thermal energy $k_{\mathrm{B}}T=\varepsilon_0=1$, the division length $r_{\mathrm{div}}=1$, and particle mass $m=1$,
so that $\tau_0=r_{\mathrm{div}}(m/k_{\mathrm{B}}T)^{1/2}=1$ is the time unit.
Here, we consider the effective temperature, including non-thermal distortion in tissues;
hence, $T$ is higher than room temperature.

For cell--cell volume exclusion, we use the short-range repulsion for $r_{ij}<r_{\mathrm {cr}}$ as in Ref.~\cite{basan2011}:
\begin{equation}\label{eq:urep}
U_{\mathrm {rep}} = \varepsilon_{\mathrm {rep}} \sum_{i,j} \Big[\big(\frac{r_{\mathrm cr}}{r_{ij}}\big)^4 + \frac{4r_{ij}}{r_{\mathrm cr}} - 5\Big]\Theta(r_{\mathrm {cr}}-r_{ij}),
\end{equation}
where $\Theta(r)$ denotes the unit step function.
In this study, $\varepsilon_{\mathrm {rep}}=1$ and $r_{\mathrm cr}=1.2$ are used.

\begin{figure*}
\includegraphics[width=16cm]{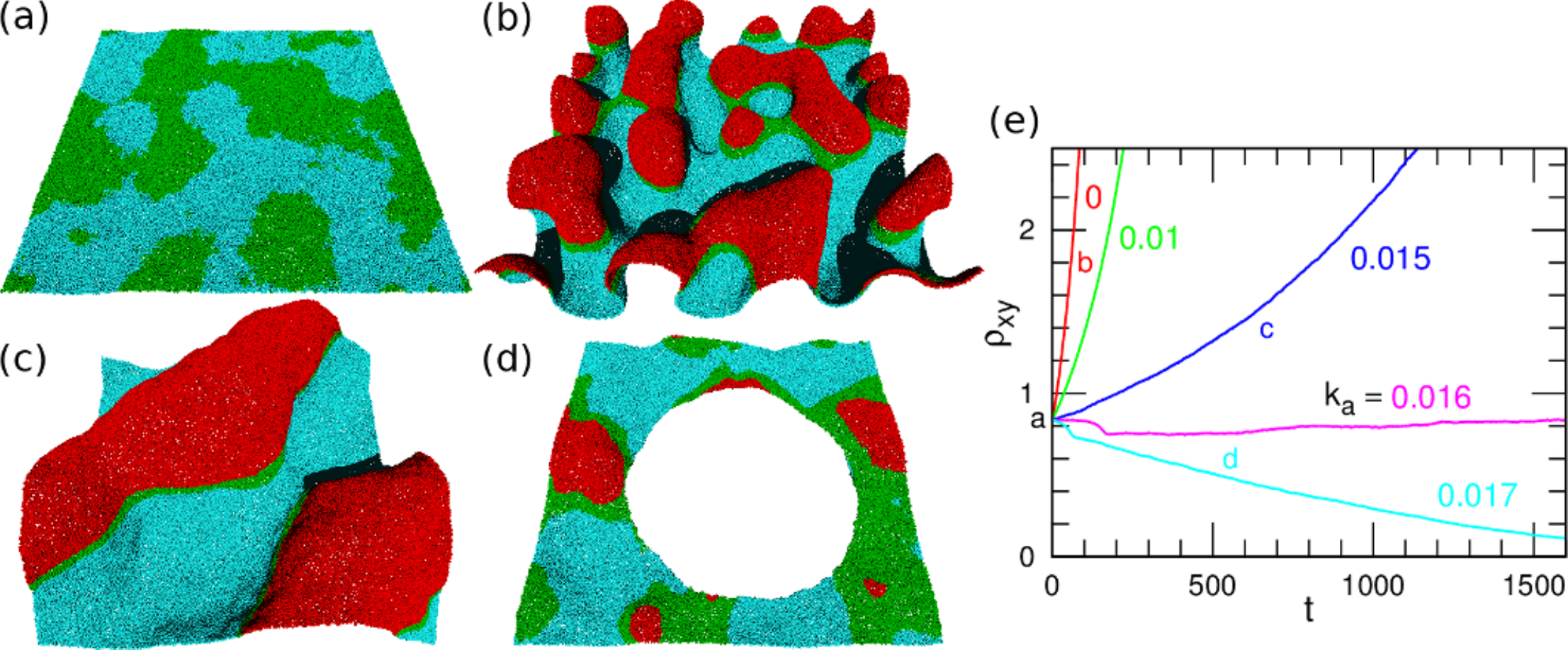}
\caption{
Tissue growth in the absence of substrate at $f_{\mathrm{div}}=4$.
(a)--(d) Snapshots. Cyan (light gray), green (medium gray), and red (dark gray) spheres represent cell particles at $z_i-z_{\mathrm{G}}\le 0$, $0<z_i-z_{\mathrm{G}}\le 5$, and $z_i-z_{\mathrm{G}}> 5$, respectively, where $z_{\mathrm{G}}$ is the $z$-component of the center of mass.
(a) Initial state. A tensionless membrane is equilibrated with no cell division and death at the cell density $\rho_{\mathrm{xy}}=0.84$ projected on the $xy$ plane.
(b) Wrinkled tissue at $\rho_{\mathrm{xy}}=1.8$ for $k_{\mathrm{a}}=0$.
(c) Buckled tissue at  $\rho_{\mathrm{xy}}=1.4$ for $k_{\mathrm{a}}=0.015$.
(d) Tissue with a pore at  $\rho_{\mathrm{xy}}=0.6$ for $k_{\mathrm{a}}=0.017$.
(e) Time evolution of $\rho_{\mathrm{xy}}$ at $k_{\mathrm{a}}=0$, $0.01$, $0.015$, $0.016$, and $0.017$ (from top to bottom).
The letters (a,b,c,d) indicate the corresponding data points of the snapshots in (a--d).
}
\label{fig:free}
\end{figure*}

To construct a one-layer tissue sheet,
we add attractive and curvature potentials ($U_{\mathrm {att}}$ and $U_{\rm {\alpha}}$) 
developed for a meshless membrane model~\cite{nogu06}.
The potential $U_{\mathrm {att}}$ is a function of 
 the local particle density $\rho_i$ as
\begin{eqnarray}\label{eq:uatt}
U_{\rm {att}} &=& \varepsilon_{\mathrm {att}} \sum_{i} \frac{1}{4}\ln[1+\exp\{-4(\rho_i-\rho^*)\}]- C,  \\ \label{eq:rho}
\rho_i &=& \sum_{j\ne i} \exp\bigg[A\Big(1+\frac{1}{(r_{ij}/r_{\mathrm {ca}})^{12} -1}\Big)\bigg]\Theta(r_{\mathrm {ca}}-r_{ij}),\ \ \ 
\end{eqnarray}
where $C= (1/4)\ln\{1+\exp(4\rho^*)\}$ and
 $A=\ln(2) \{(r_{\rm {\mathrm {ca}}}/r_{\mathrm {half}})^{12}-1\}$.
The summations in Eq.~(\ref{eq:rho}) and in the curvature potential $U_{\rm {\alpha}}$ 
are taken for all other particles, including the other particle in the same cell.
Here, $\rho_{i}$ denotes the number of particles in a 
sphere whose radius is approximately $r_{\rm {half}}$. 
This multibody potential acts as  a pair potential 
$U_{\rm {att}}\simeq -\rho_i$ with a cutoff at $\rho_i\simeq \rho^*$, 
which can stabilize
the fluid phase of 2D assembly over a wide range of parameter sets~\cite{nogu06}.
The curvature potential is given by 
\begin{eqnarray}\label{eq:ucv}
U_{\rm {\alpha}} &=& \varepsilon_{\alpha} \sum_i  \alpha_{\rm {pl}}({\bf r}_{i}), \\
\alpha_{\rm {pl}} &=& \frac{9\lambda_1\lambda_2\lambda_3}{(\lambda_1+\lambda_2+\lambda_3)
    (\lambda_1\lambda_2+\lambda_2\lambda_3+\lambda_3\lambda_1)},
\end{eqnarray}
where $\lambda_1$, $\lambda_2$, and $\lambda_3$ are
the eigenvalues of the weighted gyration tensor,
$a_{\alpha\beta}= \sum_j (\alpha_{j}-\alpha_{\rm Gw})
(\beta_{j}-\beta_{\rm Gw})w_{\rm {mls}}(r_{ij})$,
where $\alpha, \beta=x,y,z$ and ${\bf r}_{\rm Gw}=\sum_j {\bf r}_{j}w_{\rm {mls}}(r_{ij})/\sum_j w_{\rm {mls}}(r_{ij})$.
The shape parameter aplanarity $\alpha_{\rm {pl}}$ 
represents the degree of deviation from a plane and is proportional to $\lambda_1$
 for $\lambda_1 \ll \lambda_2, \lambda_3$~\cite{nogu06}.
A Gaussian function with $C^{\infty}$ cutoff~\cite{nogu06} 
is employed as a weight function 
\begin{equation}
w_{\rm {mls}}(r_{ij})=
\exp \Big[\frac{(r_{ij}/r_{\rm {ga}})^2}{(r_{ij}/r_{\rm {cg}})^{12} -1}\Big]\Theta(r_{\rm {cg}}-r_{ij}).
\end{equation}
In this study, $\varepsilon_{\mathrm {att}}=2$, $\rho^*=15$, $r_{\mathrm {ca}}=2.1$, $r_{\mathrm {half}}=1.8$, 
$\varepsilon_{\alpha}=10$, and $r_{\rm {ga}}=r_{\rm {cg}}/2 = 3$ are used.

To represent the adhesion of the tissue to a flat substrate,
we use the Lennard-Jones (LJ) potential,
\begin{equation}\label{eq:uadd}
U_{\mathrm {LJ}}=\sum_i 4\varepsilon_{\mathrm {wall}}\Big[\Big(\frac{r_{\mathrm{div}}}{z_{i}}\Big)^{12}- \Big(\frac{r_{\mathrm{div}}}{z_{i}}\Big)^{6}\Big],
\end{equation}
for the particles in all cells [see Fig.~\ref{fig:cart}(b)].
The cell particles are moved by the Newton's equation with  Langevin and dissipative particle dynamics (DPD) thermostats:
\begin{eqnarray}\label{eq:motion}
m \frac{d {\mathbf{v}}_{i}}{dt} &=&
 - \frac{\partial U}{\partial {\mathbf{r}}_i} -w_{\mathrm {wall}}(z_{i}){\mathbf{v}}_{i} +  w_{\mathrm {wall}}(z_{i})^{1/2}{\xi}_{i}(t) \\ \nonumber
&&+  \sum_{j\not=i} \left\{-w_{\mathrm {tiss}}(r_{ij}){\mathbf{v}}_{ij}
   \cdot{\hat{\mathbf{r}}}_{ij} + 
      w_{\mathrm {tiss}}(r_{ij})^{1/2}{\xi}_{ij}(t)\right\}{\hat{\mathbf{r}}}_{ij}.  
\end{eqnarray}
The second and third terms are the friction and noise from the substrate, respectively,
which work near the substrate surface.
Therefore, $w_{\mathrm {wall}}(z_{i}) = \zeta_{\mathrm {wall}}(R_{\mathrm {LT}} - z_i)\Theta(R_{\mathrm {LT}} - z_i)$ is used.
The fourth and fifth terms are the DPD thermostat (Langevin thermostat to neighboring particles keeping 
the translational and angular momenta of the pair)~\cite{espa97a,groo97}
with $w_{\mathrm {tiss}}(r_{ij})=\zeta_{\mathrm {DPD}}(R_{\mathrm {DPD}} - r_{ij})\Theta(R_{\mathrm {DPD}} - r_{ij})$.
These thermostats obey the fluctuation-dissipation theorem.
In this study, $R_{\mathrm {LT}}=1.5$, $R_{\mathrm {DPD}}=2.1$, and $\zeta_{\mathrm {DPD}}=1$ are fixed, and $\zeta_{\mathrm {wall}}$ is varied to investigate the friction effects.
Note that the cell division and death do not conserve the translational and angular momenta, although the DPD thermostat conserves them.
The cell motion is numerically solved by the Shardlow's S1 splitting algorithm~\cite{shar03,nogu07a}
with $\Delta t_{\mathrm{MD}}=0.002$, and the cell division and death are performed every $\Delta t_{\mathrm{div}}=0.04$.

To generate the initial states and  calculate the thermal properties,
tissues are equilibrated in the absence of cell division and death with an additional potential
\begin{equation}\label{eq:uinit}
U_{\mathrm {init}} = \sum_{k}  \exp[20(r_{\mathrm{cl},k}-r_{\mathrm {div}})],
\end{equation}
to keep a distance of less than $r_{\mathrm {div}}$.
The tissue sheet is in the fluid phase 
and has a bending rigidity $\kappa \simeq 50\varepsilon_0$, edge line tension $\Gamma \simeq 13\varepsilon_0/r_{\mathrm{div}}$,
and membrane viscosity $\eta = 8\varepsilon_0 \tau_0 r_{\mathrm{div}}^{-2}$ (calculation details are given in \ref{sec:eq}).
The simulation units can be mapped to real scale using these quantities.
The bending rigidity and membrane viscosity of tissues depend on inter-cell adhesion energy and tissue thickness. 
The bending rigidity is estimated as $\kappa \sim 10^{-15}$ -- $10^{-11}\,$J~\cite{hann14}. 
The 3D viscosity of the aggregate of mouse embryonic cells is measured as $4\times 10^5\,$Pa$\cdot$s~\cite{marm09}.
When we consider a tissue sheet with a thickness of $10\,\mu$m, $\kappa \sim 5\times 10^{-14}\,$J, and $\eta \sim 4\,$Pa$\cdot$s$\cdot$m$^{-1}$,
the simulation units correspond to $r_{\mathrm{div}} \sim 10\,\mu$m,
$\varepsilon_0 \sim 10^{-15}\,$J,
and $\tau_0 \sim 10\,$h on a real scale.
Statistical errors are calculated from three (or more) and ten independent runs
for steady states and time evolutions, respectively.

\begin{figure*}
\includegraphics[width=16cm]{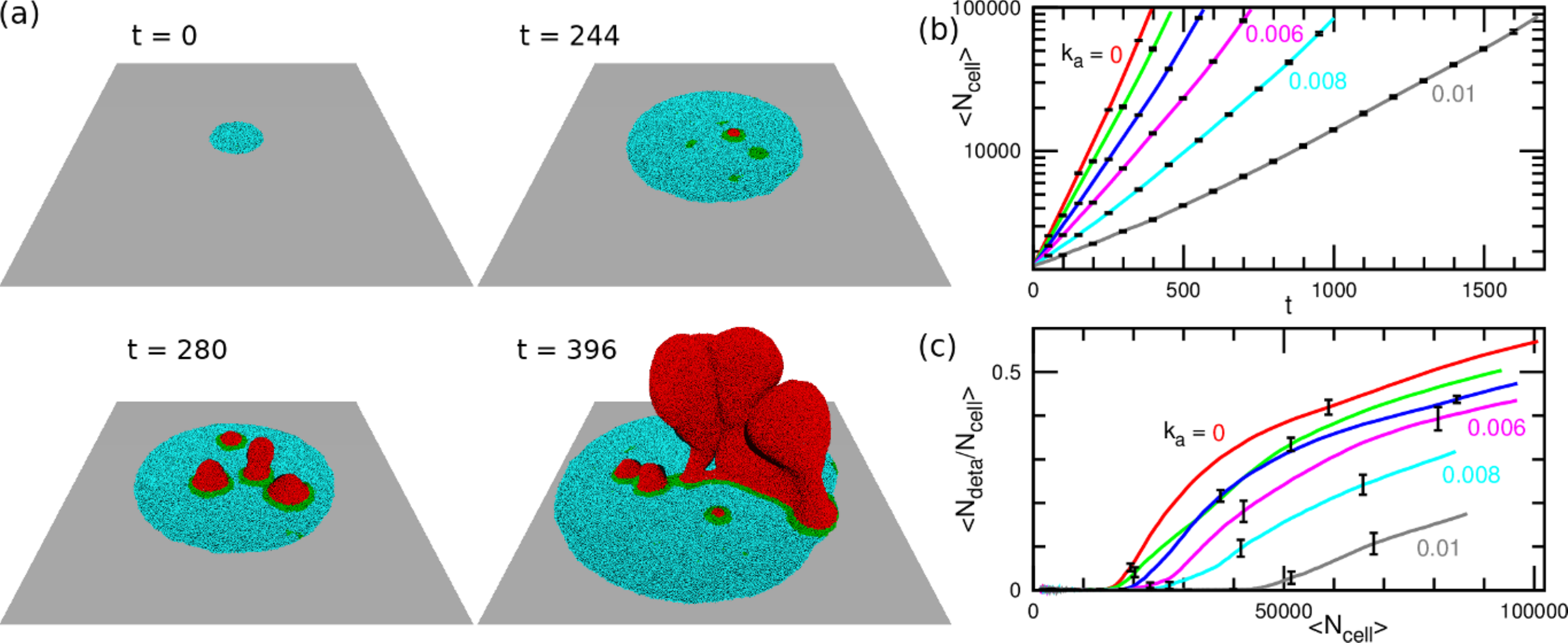}
\caption{
Tissue growth with budding on substrate at $f_{\mathrm{div}}=4$, $\varepsilon_{\mathrm{wall}}=0.1$, and $\zeta_{\mathrm{wall}}=2$.
(a) Sequential snapshots at $k_{\mathrm{a}}=0$.
Cyan (light gray), green (medium gray), and red (dark gray) spheres represent cell particles at $z_i\le 2$, $2<z_i\le 5$, and $z_i> 5$, respectively.
(b) Time evolution of mean number $\langle N_{\mathrm {cell}}\rangle$ of cells at $k_{\mathrm{a}}=0$, $0.002$, $0.004$, $0.006$, $0.008$, and $0.01$ (from left to right).
(c) Mean ratio of detached cells $\langle N_{\mathrm{deta}}/N_{\mathrm{cell}}\rangle$ as a function of $\langle N_{\mathrm {cell}}\rangle$  at $k_{\mathrm{a}}=0$, $0.002$, $0.004$, $0.006$, $0.008$, and $0.01$ (from top to bottom).
}
\label{fig:f2f4}
\end{figure*}

\section{Freely suspended tissue sheet}\label{sec:free}

Before considering the interaction with the substrate,
we investigate the dynamics of freely suspended tissues in the absence of a substrate.
The tissue is initially in a tensionless state equilibrated in the absence of cell division and death with $U_{\mathrm {init}}$
under the periodic boundary conditions at $L_x=L_y=200$ and $N_{\mathrm{cell}}=33776$ [see Fig.~\ref{fig:free}(a)].
As the cells start dividing without death ($k_{\mathrm{a}}=0$),  the tissue forms wrinkles [see Fig.~\ref{fig:free}(b) and Movie S1].
Neighboring small wrinkles fuse, but large $\Omega$-shaped wrinkles do not, since the contact of the foot regions is prevented by the convex regions.
As the death rate $k_{\mathrm{a}}$ increases, the growth becomes slower, and the number of wrinkles decreases through more frequent fusion, since the tissues can relax into a shape of lower bending energy.
Eventually, a single buckle is formed (Fig.~\ref{fig:free}(c)), since it has the lowest bending energy (see \ref{sec:eq}). This shape is called elastica~\cite{true83} that is formed in both elastic and fluid sheets~\cite{nogu11a}.
With a further increase in $k_{\mathrm{a}}$, the tissue shrinks and disappears via forming a pore, which subsequently grows [Figs.~\ref{fig:free}(d) and (e)]. 
At the threshold condition $k_{\mathrm{a}}=0.016$, the tissue forms a pore but subsequently grows and closes the pore.
This initial slow growth rate is due to the initial relaxation of $r_{\mathrm{cl},k}$ by the removal of $U_{\mathrm {init}}$.

These dynamics seem in contrast to previous simulations by the cell vertex~\cite{inou20} and cell-centered models~\cite{mimu23}.
They obtained straight or winding bumps are formed instead of buds.
We speculate that the difference comes from the suppression of the vertical growth in their simulations
owing to the harmonic constraint potential for vertical cell positions.

Here, we use a DPD thermostat, such that the viscous tissue is placed in a vapor of negligible viscosity.
Therefore, the tissue can move more easily in the vertical than lateral direction.
When we use the Langevin thermostat instead, the tissue can move equally in both directions,
so that more frequent wrinkles are formed owing to less frequent wrinkle fusion.
The viscosity ratio exerts similar effects in lipid membranes~\cite{naka18}.

\section{Growing tissue on substrate with constant division force}\label{sec:disk}

\begin{figure*}
\includegraphics[]{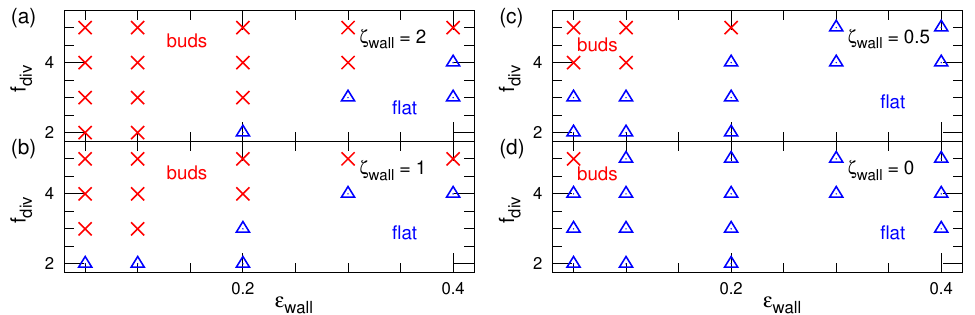}
\caption{
Dynamic phase diagrams of tissue growth on substrate with constant $f_{\mathrm{div}}$ at $k_{\mathrm{a}}=0$.
(a) $\zeta_{\mathrm{wall}}=2$. (b) $\zeta_{\mathrm{wall}}=1$. (c) $\zeta_{\mathrm{wall}}=0.5$. 
(d) $\zeta_{\mathrm{wall}}=0$. 
The red crosses and blue triangles represent budded tissues and flat tissues without budding, respectively.
}
\label{fig:pdfe}
\end{figure*}

We investigate the tissue growth on a substrate with constant $f_{\mathrm{div}}$;
the cell environment, such as the nutrient concentration, is assumed to be uniform, including far places from the substrate.
The initial state is a disk-shaped tissue equilibrated in the absence of cell division and death 
with $U_{\mathrm {init}}$ at $N_{\mathrm {cell}}=1600$.
The tissues spread on the substrate and subsequently form buds [see Fig.~\ref{fig:f2f4}(a) and Movie S2].
The number of cells $N_{\mathrm {cell}}$ increases exponentially. 
As $k_{\mathrm{a}}$ increases, the growth speed becomes lower, and the budding begins at larger tissues [see Figs.~\ref{fig:f2f4}(b) and (c)].
For quantification, we consider cells at $z_{\mathrm{cc},k}\ge 2$ to be detached from the substrate,
where $z_{\mathrm{cc},k}$ is the height of cell center: $z_{\mathrm{cc},k} = (z_i +z_j)/2$ where the heights $z_i$ and $z_j$ of the two particles in the $k$-th cell.

Figure \ref{fig:pdfe} shows the dynamic phase diagrams for $\varepsilon_{\mathrm{wall}}$ vs. $f_{\mathrm{div}}$ at four $\zeta_{\mathrm{wall}}$ values.
The buds are formed under weaker adhesion (low $\varepsilon_{\mathrm{wall}}$) and/or faster growth (high $f_{\mathrm{div}}$) 
with stronger substrate friction $\zeta_{\mathrm{wall}}$.
We consider that a tissue forms buds when the ratio of detached cells is more than  $0.1$ at $N_{\mathrm {cell}}=100~000$
for $\varepsilon_{\mathrm{wall}}\ge 0.1$ and $0.25$ for $\varepsilon_{\mathrm{wall}}=0.05$.
For a weak adhesion of $\varepsilon_{\mathrm{wall}}=0.05$, $10$\% of cells are occasionally detached under thermal fluctuations; 
eventually, the growing tissues can be entirely detached from the substrate.

\begin{figure}
\includegraphics[]{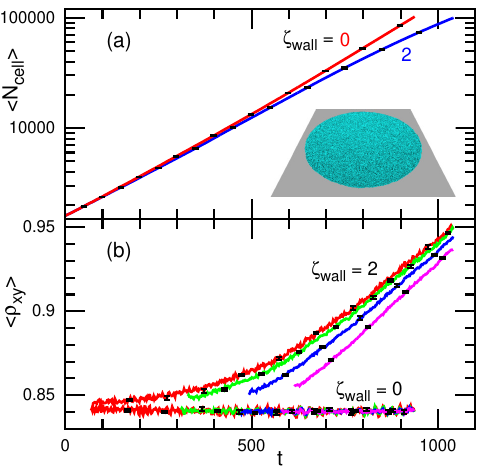}
\caption{
Tissue growth without budding on substrate with  $\zeta_{\mathrm{wall}}=0$ and $2$
at $f_{\mathrm{div}}=3$, $\varepsilon_{\mathrm{wall}}=0.4$, and $k_{\mathrm{a}}=0$.
(a) Time evolution of mean number of cells $\langle N_{\mathrm {cell}}\rangle$.
(b) Time evolution of mean local density $\langle \rho_{xy}\rangle$ on projected on the $xy$ plane.
The red, green, blue, and magenta represent the data for $r_{\mathrm{2D}}< 20$, $20\le r_{\mathrm{2D}} <40$,  $40\le r_{\mathrm{2D}} <60$, and  $60\le r_{\mathrm{2D}} <80$, respectively (from left to right), where $r_{\mathrm{2D}}$ is the distance from the center of tissues on the $xy$ plane. 
The inset in (a) shows a snapshot at $t=1000$ for $\zeta_{\mathrm{wall}}=2$.
}
\label{fig:flat}
\end{figure}

For no substrate friction ($\zeta_{\mathrm{wall}}=0$), 
the budding does not occur except for the weakest adhesion, $\varepsilon_{\mathrm{wall}}=0.05$ [see Fig.~\ref{fig:pdfe}(d)].
This suggests that friction plays an essential role in the budding.
To clarify it, we calculate the local density $\rho_{xy}$ projected on the $xy$ plane in tissues spreading on the substrate without budding, as shown in Fig.~\ref{fig:flat}.
The center region of tissue,  $\rho_{xy}$ increases with tissue growth at $\zeta_{\mathrm{wall}}=2$, 
whereas $\rho_{xy}$ remains constant at $\zeta_{\mathrm{wall}}=0$.
Hence, the cells can move sufficiently to maintain their preferred density at $\zeta_{\mathrm{wall}}=0$ during the exponential growth [see the red line in Fig.~\ref{fig:flat}(a)]. Conversely, friction slows relaxation, resulting in high density and high stress in the center region,
since friction increases under increasing growth speed with time (the velocity of tissue radius $\mathrm{d}R/\mathrm{d}t\propto \exp(at/2)$ for $N_{\mathrm {cell}}=\pi R^2\rho_{xy}=N_0\exp(at)$ with constant $\rho_{xy}$).
Budding occurs,
when this high stress overcomes substrate adhesion.
Note that a similar retardation in growth speed with increasing friction was reported in an overdamped Langevin simulation using a 2D vertex model~\cite{guer23}. Therefore, this mechanism is likely generic for tissue growth.

\section{Growing tissue on substrate with height-dependent division force}\label{sec:steady}

\begin{figure*}
\includegraphics[width=16cm]{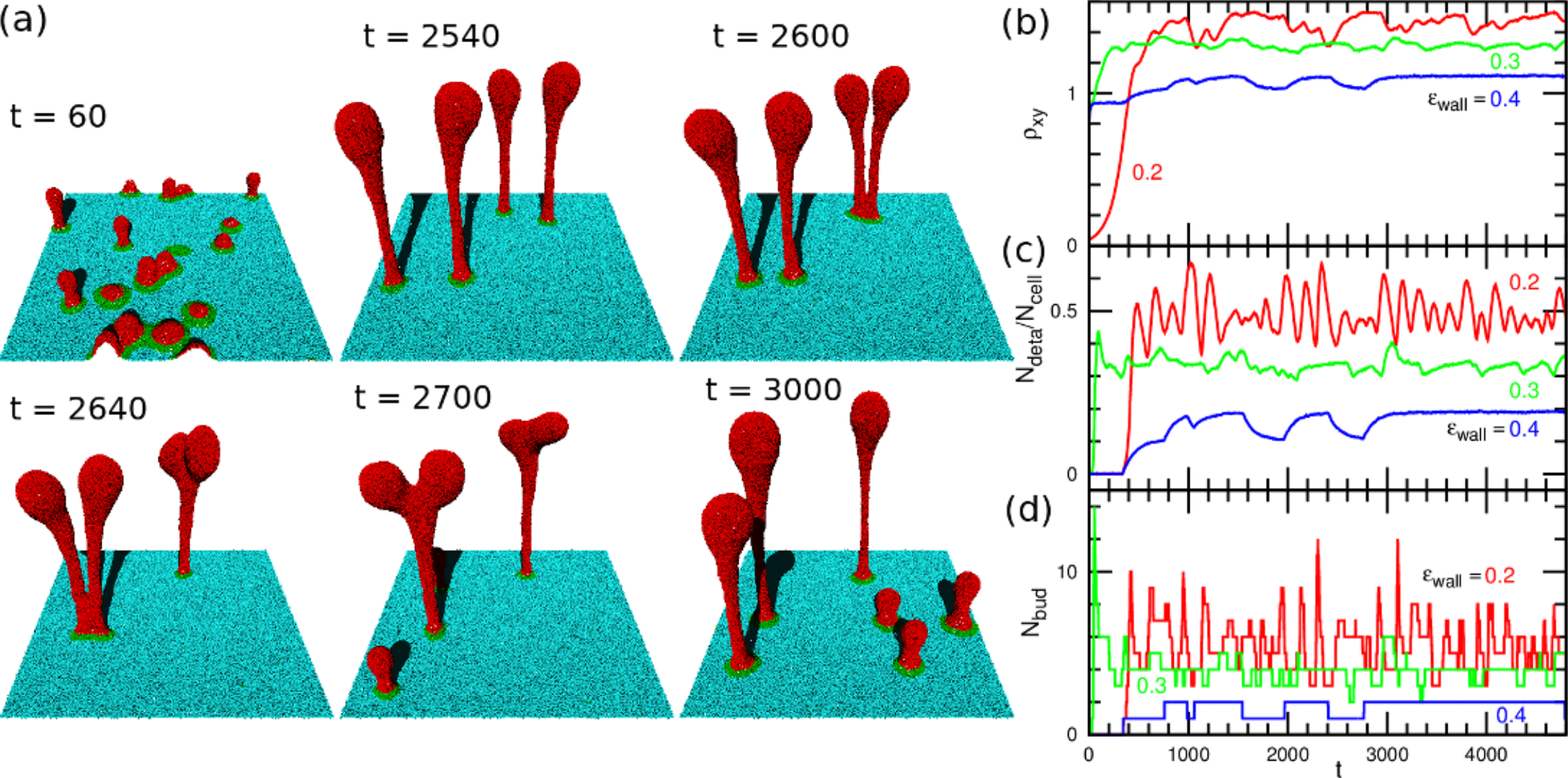}
\caption{
Tissue growth with budding on substrate with non-uniform $f_{\mathrm{div}}$ 
at $f_{\mathrm{da}}=5$, $k_{\mathrm{a}}=0.01$, and $\zeta_{\mathrm{wall}}=2$.
(a) Sequential snapshots at $\varepsilon_{\mathrm{wall}}=0.3$.
(b)--(d) Time evolution at $\varepsilon_{\mathrm{wall}}=0.2$, $0.3$, and $0.4$.
(b) Cell density $\rho_{xy}$ projected on the $xy$ plane.
(c) Ratio of detached cells $N_{\mathrm{deta}}/N_{\mathrm{cell}}$.
(d) Number of buds $N_{\mathrm {bud}}$.
}
\label{fig:hf5a10}
\end{figure*}

\begin{figure*}
\includegraphics[width=16cm]{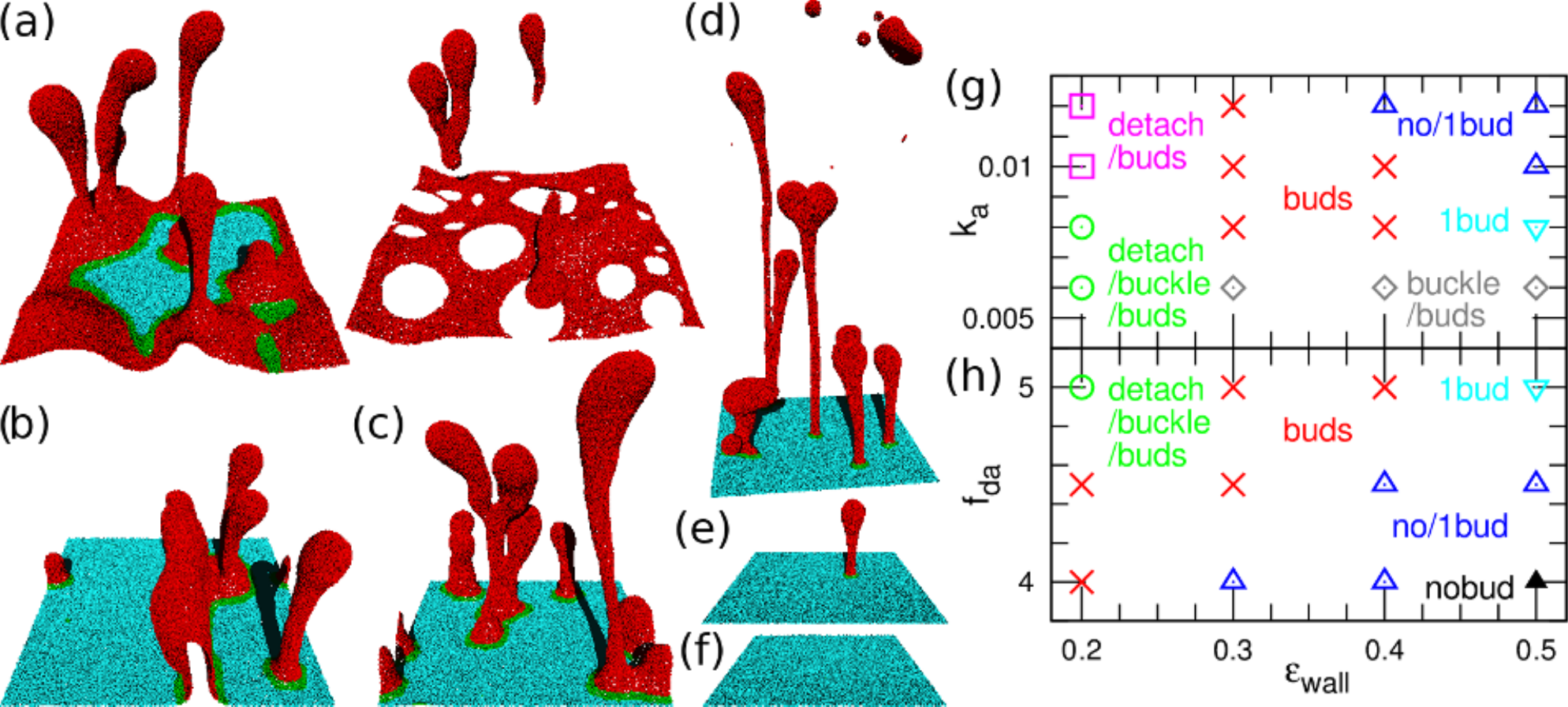}
\caption{
Tissue growth on substrate with non-uniform $f_{\mathrm{div}}$ at $\zeta_{\mathrm{wall}}=2$.
(a)--(c) Snapshots at $f_{\mathrm{da}}=5$, $\varepsilon_{\mathrm{wall}}=0.2$, and $k_{\mathrm{a}}=0.008$. 
(a) Sequential snapshots of tissue detachment at $t=1420$ and $1500$.
(b) Buckling with buds.
(c) Budding.
(d) Snapshot of budding with vesicle pinch-off at $f_{\mathrm{da}}=5$, $\varepsilon_{\mathrm{wall}}=0.3$, and $k_{\mathrm{a}}=0.006$. 
(e)--(f) Snapshots at $f_{\mathrm{da}}=5$, $\varepsilon_{\mathrm{wall}}=0.4$, and $k_{\mathrm{a}}=0.012$. 
(g)--(h) Dynamic phase diagrams of tissue growth for (g)  $f_{\mathrm{da}}=5$,  and (h) $k_{\mathrm{a}}=0.008$.
The red crosses represent the formation of multiple buds.
The blue upward-pointing triangles represent the coexistence of flat tissues without buds and with one bud.
The light blue downward-pointing triangles and black filled triangles represent the flat tissues with one bud and with no buds, respectively.
The green circles represent the coexistence of tissue detachment, buckling, and multiple budding.
The magenta squares represent the coexistence of
tissue detachment and multiple budding.
The gray diamonds represent the coexistence of
tissue buckling and multiple budding.
}
\label{fig:hpd}
\end{figure*}

In cultured-tissue experiments, nutrients are typically supplied from a gel substrate, and in epithelia, growth factors come from the stroma below.
Hence, tissues may not be able to grow at a place far from the substrate because of a lack of nutrients.
The apoptosis rate also depends on environmental conditions.
In anoikis, the cell detachment from the extracellular matrix induces apoptosis~\cite{fris01,gros02}.
Under the former and latter conditions, the division force $f_{\mathrm{div}}$ and death rate $k_{\mathrm{a}}$ are functions of the cell height, respectively. 
For simplicity, we set $f_{\mathrm{div}}$ as a function of the height $z_{\mathrm{cc},k}$ of the cell center while keeping $k_{\mathrm{a}}$ constant:
\begin{equation}
f_{\mathrm{div}} = \frac{f_{\mathrm{da}}}{ 1 + \exp[2(z_{\mathrm{cc},k} -3)]}.
\end{equation}
The force strength decreases from $f_{\mathrm{div}} \simeq f_{\mathrm{da}}$ to null at $z_{\mathrm{cc},k}\sim 3$ [see Fig.~\ref{fig:cart}(c)].
We use $f_{\mathrm{da}}= 4$ to $5$, $k_{\mathrm{a}}=0.006$ to $0.012$, and $\zeta_{\mathrm{wall}}=2$ at $L_x=L_y=200$.
The tissue grows on the substrate but shrinks at $z_{\mathrm{cc},k}\gtrsim 3$.

Figure~\ref{fig:hf5a10} shows examples of tissue dynamics during steady bud evolution.
Tubular bud elongation is arrested at a certain height, 
neighboring buds move closer and subsequently fuse into one bud, and
 new buds are formed in an open space [see Fig.~\ref{fig:hf5a10}(a) and Movie S3].
Bud formation and fusion occur repeatedly, so that
the cell density $\rho_{xy}$, the ratio of the detached cell $N_{\mathrm{deta}}/N_{\mathrm{cell}}$, and number of buds $N_{\mathrm {bud}}$
fluctuate around certain values [see Figs.~\ref{fig:hf5a10}(b)--(d)].
The number of buds is counted as the number of cell clusters at $5<z_{\mathrm{cc},k} <8$ (foot of the red regions in snapshots) 
to exclude cells pinched off from the tissue.
Bud fusion is caused by the membrane-mediated attraction, which reduces the bending energy of the adhered tissue (cyan and green region in snapshots).
A similar membrane-mediated attraction has been observed in tubules generated by curvature-inducing proteins~\cite{nogu16,nogu22b}.

The dynamic phase diagrams are shown in Figs.~\ref{fig:hpd}(g) and (h).
Similar to the case of constant $f_{\mathrm{div}}$, the mean number of buds decreases with decreasing $f_{\mathrm{da}}$ and increasing $k_{\mathrm{a}}$, 
and eventually, no buds are formed from the flat tissue at the region of upward-pointing triangles in Figs.~\ref{fig:hpd}(f)--(h).
However, when a budded state is set as the initial conformation, a single bud remains in a steady state after bud fusion [blue-triangle points in Fig.~\ref{fig:hpd}(e)].
Hence, two types of final states are formed depending on the initial states.

\begin{figure*}
\includegraphics[]{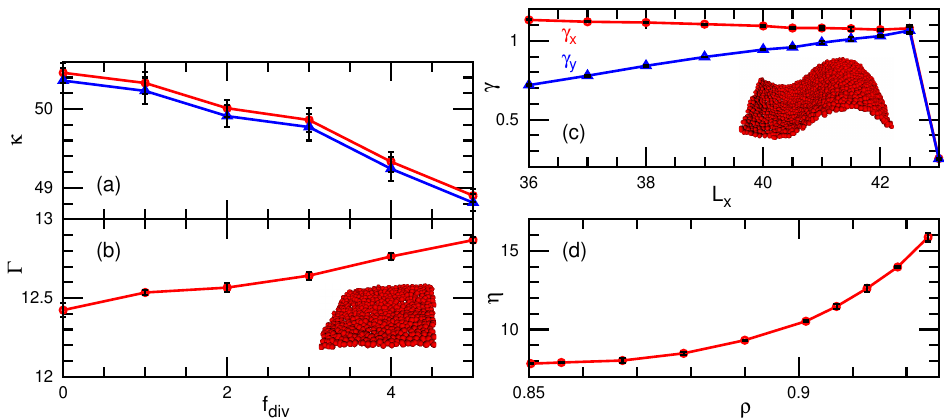}
\caption{
Tissue membrane properties in thermal equilibrium (no cell division and death).
(a) Bending rigidity $\kappa$ calculated from buckling of membranes.
The red and blue lines are obtained using Eqs.~(\ref{eq:cv1}) and (\ref{eq:cv2}), respectively.
(b) Edge line tension $\Gamma$ calculated from membrane strips.
The inset shows a snapshot at $f_{\mathrm{div}}=5$ and $L_x=20$.
(c) Stresses $\gamma_x$ and $\gamma_y$ in the $x$ and $y$ directions, respectively, 
in buckled membranes at $f_{\mathrm{div}}=5$.
The inset shows a snapshot at $L_x=36$.
(d) Viscosity $\eta$ dependence on cell density $\rho$ at $f_{\mathrm{div}}=5$ in the 2D space.
}
\label{fig:eq}
\end{figure*}

At a low death rate of $k_{\mathrm{a}}=0.006$, vesicle division often occurs in long tubular buds 
via tube pinch-off (like droplet formation by Plateau-Rayleigh instability~\cite{dege03,utad05}) 
and tube branching during bud fusion [see Fig.~\ref{fig:hpd}(d)].
For the weak adhesion of $\varepsilon_{\mathrm{wall}}=0.2$, entire tissues are often detached from the substrate [see Fig.~\ref{fig:hpd}(a) and Movie S4]. In particular, the entire detachment occurs more frequently when a flat tissue [as in Fig.~\ref{fig:hpd}(f)] is used as the initial state. In addition, a straight buckled bump [like the buckling of equilibrated membrane shown in Fig.~\ref{fig:eq}(c)] is occasionally formed with the weak adhesion and/or low death rate [see Fig.~\ref{fig:hpd}(b) and Movie S5].
Tubular buds move towards this bump and fuse into it. When two buckled bumps are set as the initial state, two bumps fuse into a single one (see  Movie S6),
such that the bumps also have the membrane-mediated attraction as tubular buds.
Note that with a buckled state set as the initial state, the buckle remains in the most region of red crosses in the diagrams. 
Thus, different tissue shapes can be obtained depending on the initial state.

\begin{figure*}
\includegraphics[width=16cm]{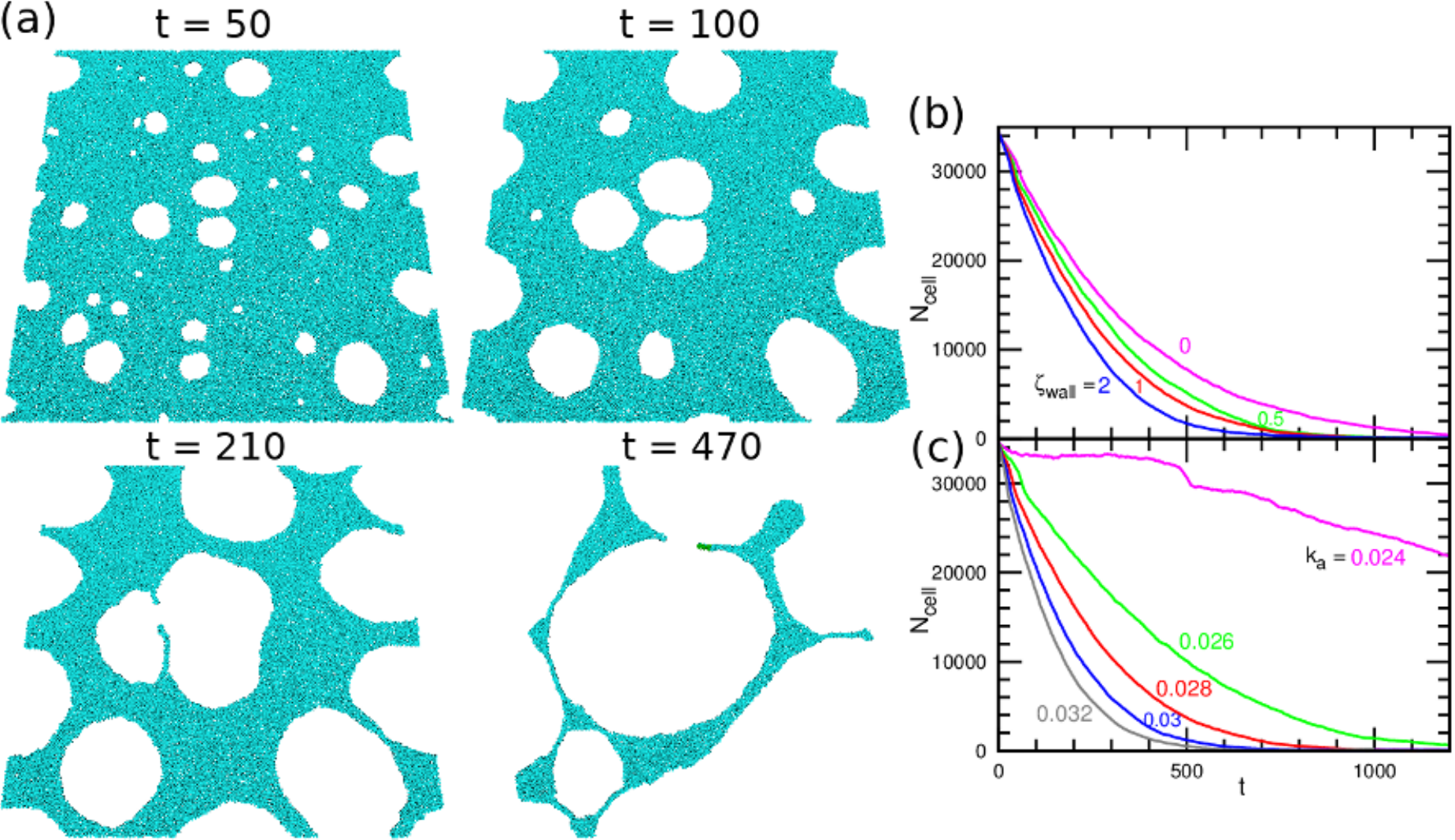}
\caption{
Pore formation in tissues on substrate at $f_{\mathrm{div}}=5$ and $\varepsilon_{\mathrm{wall}}=0.4$.
(a) Sequential snapshots at $\zeta_{\mathrm{wall}}=1$ and $k_{\mathrm{a}}=0.028$.
(b)--(c) Time evolution of number of cells $N_{\mathrm {cell}}$.
(b) $\zeta_{\mathrm{wall}}=0$, $0.5$, $1$, and $2$ (from top to bottom) at $k_{\mathrm{a}}=0.028$.
(c) $k_{\mathrm{a}}=0.024$, $0.026$, $0.028$, $0.03$, and $0.032$ (from top to bottom) at $\zeta_{\mathrm{wall}}=1$.
The red curves in (b) and (c) correspond to the data shown as snapshots in (a).
}
\label{fig:pore}
\end{figure*}

\section{Pore formation in flat tissues on substrate}\label{sec:pore}

The formation of pores (or wounds) in tissue sheets and their healing are important for tissue maintenance~\cite{tetl19,prak21,xu22,sona23,mcev22}.  
Recently, Lv et al.~\cite{lv24} observed the coalescence of pores in epithelioid tissues on a hydrogel bed and simulated it using a cell-vertex-based model.
However, the effects of physical conditions such as friction were not investigated. 
Here, we simulate the pore formation under various conditions of friction and cell death.

The temporal evolution of tissue pores is shown in Fig.~\ref{fig:pore} and Movie S7. The initial state is a flat tissue relaxed at $f_{\mathrm{div}}=5$, $\varepsilon_{\mathrm{wall}}=0.4$, and $k_{\mathrm{a}}=0.02$. The other parameters are the same as those described in Sec.~\ref{sec:steady}.
First, many pores simultaneously open, and subsequently, 
larger pores grow; however, small pores shrink through Ostwald ripening~\cite{voor85,kaba01,wata14} 
(compare the top two snapshots in Fig.~\ref{fig:pore}(a)). Coalescence of the pores also occurs (see the bottom two snapshots in Fig.~\ref{fig:pore}(a) and Movie S7).
The Ostwald ripening more frequently occurs at lower friction levels. At $\zeta_{\mathrm{wall}}=0$, a single pore remains for a long period but the others disappear via pore shrinkage, similar to that for freely suspended tissues (see Fig.~\ref{fig:free}(d)). This is because the tissues can move easily in the lateral directions in the absence of substrate friction. As the friction increases, coalescence occurs more frequently than the Ostwald ripening.  
In contrast, the tissue shrinkage is only slightly dependent on the friction, as shown in Fig.~\ref{fig:pore}(b). With an increasing death rate $k_{\mathrm{a}}$, the tissues exhibit faster shrinkage (see Fig.~\ref{fig:pore}(c)) and more frequent pore coalescence. Thus, the pore growth dynamics are strongly dependent on substrate friction and pore growth rate.

\section{Conclusions and outlook}\label{sec:sum}

By combining the two-particle growth model with a meshless membrane model, we constructed a particle-based simulation technique as a minimum model for tissue sheet growth and shrinkage.
Using this combined model, we have studied the growth and shrinkage of tissue sheets.
Freely suspended sheets form wrinkles for rapid tissue growth. 
The wrinkling number decreases with decreasing growth rate.
On the substrate, the growing tissue forms buds when the growth rate is too high to maintain the 2D density.
When the induced stress overcomes the adhesion strength, local tissue detachment results in the formation of cylindrical buds.
With stronger friction and/or weaker adhesion to the substrate, buds are formed more frequently.
Neighboring buds fuse through a membrane-mediated attraction. 
The detachment of entire tissues and straight buckled bump formation also occur under weak adhesion conditions. Moreover, tissue pores (wounds) grow through Ostwald ripening and coalescence. The coalesce occurs more frequently with higher friction and/or death rates.

Our simulations revealed that adhesion and friction with the substrate are important factors in controlling the detachment of growing tissues.
Experimentally, the adhesion energy to a substrate can be changed by varying the protein concentrations for ligand-receptor binding and surface charges. These variations and surface roughness change the substrate friction. Therefore, our findings can be examined experimentally.
Similar detachment dynamics can be observed in epithelial tissues from different tissues with weakened adhesion by wounds and diseases.

The skin and brain cortex typically consist of several layers of tissue sheets with different elasticities.
It is known that such a difference can generate wrinkling~\cite{schw09,li12}. The proposed model can be extended to multiple sheets with different rigidities. The growth dynamics of multi-layer tissues is one direction for further study.

\begin{appendix}

\section{Physical properties of tissues}\label{sec:eq}

The tissue membrane properties are calculated in the absence of cell division or death.
The bending rigidity $\kappa$ of membranes can be calculated using several methods~\cite{dese09,shib11}.
Here, we calculated $\kappa$ from the buckling of membranes~\cite{nogu11a,hu13a},
since its condition (positive surface stress) is closer to the present simulation conditions.
A flat membrane with $N_{\mathrm{cell}}=800$ set along the $xy$ plane 
 is pushed along the $x$ axis by changing $L_x$ while keeping $L_y=22$.
After the membrane buckling,
the stresses $\gamma_x$ and $\gamma_y$ along the $x$ and $y$ axes are different owing to the increase in the bending energy.
The bending rigidity $\kappa$ can be estimated from $\gamma_x$ as
\begin{eqnarray}\label{eq:cv1}
\gamma_x &=& \frac{\pi^2\kappa}{4L^2}\Big(5 - \frac{L_x}{L}\Big)^2 + O\bigg(\Big(1-\frac{L_x}{L}\Big)^2\bigg), \\ \label{eq:cv2}
 &=& \frac{4\pi^2\kappa}{L^2}\bigg[1+ \frac{1}{2}\Big(1 - \frac{L_x}{L}\Big) + \frac{9}{32}\Big(1 - \frac{L_x}{L}\Big)^2   \nonumber \\
&& +\frac{21}{128}\Big(1 - \frac{L_x}{L}\Big)^3\bigg] + O\bigg(\Big(1-\frac{L_x}{L}\Big)^4\bigg), \hspace{1cm}
\end{eqnarray}
in the leading-~\cite{nogu11a} and third-order~\cite{hu13a} approximations, respectively,
where the tensionless membrane area is $A=LL_y$.
Both approximations can give excellent fits to the $\gamma_x$ curve in Fig.~\ref{fig:eq}(c), 
and $\kappa$ is estimated as shown in Fig.~\ref{fig:eq}(a).

The edge line tension $\Gamma$ is calculated from the stress of a membrane strip with $N_{\mathrm{cell}}=400$.
Since the edge energy per length is given by $\Gamma$, the edge tension is expressed as $\Gamma = -P_{xx}L_yL_z/2$,
where $P_{xx}$ is the $x$ component of the pressure calculated by the virial tensor~\cite{tolp04,shib11}.
The obtained values of $\kappa$ and  $\Gamma$ exhibit only a weak dependence on $f_{\mathrm{div}}$ [see Figs.~\ref{fig:eq}(a) and (b)],
so that we consider $\kappa \simeq 50$ and $\Gamma \simeq 13$ 
even in the presence of cell division and death.
These values are sufficiently large to maintain a flat tissue sheet with a minimum edge length (circular edge for disk-shaped tissue).

The membrane viscosity $\eta$ of the 2D tissue (i.e., $\varepsilon_{\mathrm{wall}}\to\infty$) is calculated from the tissue stress under simple shear flow at $L_x=40$ and $L_y=10$~\cite{nogu07a}.
The viscosity $\eta$ increases with increasing cell density $\rho$ [see Fig.~\ref{fig:eq}(d)].
Since the typical density is $0.85$ in our simulations (see Fig.~\ref{fig:flat}), 
we consider $\eta = 8$ as the typical viscosity of the tissue.

\end{appendix}

\begin{acknowledgments}
This work was supported by JSPS KAKENHI Grant Number JP24K06973.
\end{acknowledgments}

%\bibliographystyle{apsrev4-1}
%\bibliography{ref0}

\begin{thebibliography}{77}%
\makeatletter
\providecommand \@ifxundefined [1]{%
 \@ifx{#1\undefined}
}%
\providecommand \@ifnum [1]{%
 \ifnum #1\expandafter \@firstoftwo
 \else \expandafter \@secondoftwo
 \fi
}%
\providecommand \@ifx [1]{%
 \ifx #1\expandafter \@firstoftwo
 \else \expandafter \@secondoftwo
 \fi
}%
\providecommand \natexlab [1]{#1}%
\providecommand \enquote  [1]{``#1''}%
\providecommand \bibnamefont  [1]{#1}%
\providecommand \bibfnamefont [1]{#1}%
\providecommand \citenamefont [1]{#1}%
\providecommand \href@noop [0]{\@secondoftwo}%
\providecommand \href [0]{\begingroup \@sanitize@url \@href}%
\providecommand \@href[1]{\@@startlink{#1}\@@href}%
\providecommand \@@href[1]{\endgroup#1\@@endlink}%
\providecommand \@sanitize@url [0]{\catcode `\\12\catcode `\$12\catcode
  `\&12\catcode `\#12\catcode `\^12\catcode `\_12\catcode `\%12\relax}%
\providecommand \@@startlink[1]{}%
\providecommand \@@endlink[0]{}%
\providecommand \url  [0]{\begingroup\@sanitize@url \@url }%
\providecommand \@url [1]{\endgroup\@href {#1}{\urlprefix }}%
\providecommand \urlprefix  [0]{URL }%
\providecommand \Eprint [0]{\href }%
\providecommand \doibase [0]{http://dx.doi.org/}%
\providecommand \selectlanguage [0]{\@gobble}%
\providecommand \bibinfo  [0]{\@secondoftwo}%
\providecommand \bibfield  [0]{\@secondoftwo}%
\providecommand \translation [1]{[#1]}%
\providecommand \BibitemOpen [0]{}%
\providecommand \bibitemStop [0]{}%
\providecommand \bibitemNoStop [0]{.\EOS\space}%
\providecommand \EOS [0]{\spacefactor3000\relax}%
\providecommand \BibitemShut  [1]{\csname bibitem#1\endcsname}%
\let\auto@bib@innerbib\@empty
%</preamble>
\bibitem [{\citenamefont {Beumer}\ and\ \citenamefont
  {Clevers}(2021)}]{beum21}%
  \BibitemOpen
  \bibfield  {author} {\bibinfo {author} {\bibfnamefont {J.}~\bibnamefont
  {Beumer}}\ and\ \bibinfo {author} {\bibfnamefont {H.}~\bibnamefont
  {Clevers}},\ }\href {\doibase 10.1038/s41580-020-0278-0} {\bibfield
  {journal} {\bibinfo  {journal} {Nat. Rev. Mol. Cell Biol.}\ }\textbf
  {\bibinfo {volume} {22}},\ \bibinfo {pages} {39} (\bibinfo {year}
  {2021})}\BibitemShut {NoStop}%
\bibitem [{\citenamefont {Shyer}\ \emph {et~al.}(2013)\citenamefont {Shyer},
  \citenamefont {Tallinen}, \citenamefont {Nerurkar}, \citenamefont {Wei},
  \citenamefont {Gil}, \citenamefont {Kaplan}, \citenamefont {Tabin},\ and\
  \citenamefont {Mahadevan}}]{shye13}%
  \BibitemOpen
  \bibfield  {author} {\bibinfo {author} {\bibfnamefont {A.~E.}\ \bibnamefont
  {Shyer}}, \bibinfo {author} {\bibfnamefont {T.}~\bibnamefont {Tallinen}},
  \bibinfo {author} {\bibfnamefont {N.~L.}\ \bibnamefont {Nerurkar}}, \bibinfo
  {author} {\bibfnamefont {Z.}~\bibnamefont {Wei}}, \bibinfo {author}
  {\bibfnamefont {E.~S.}\ \bibnamefont {Gil}}, \bibinfo {author} {\bibfnamefont
  {D.~L.}\ \bibnamefont {Kaplan}}, \bibinfo {author} {\bibfnamefont {C.~J.}\
  \bibnamefont {Tabin}}, \ and\ \bibinfo {author} {\bibfnamefont
  {L.}~\bibnamefont {Mahadevan}},\ }\href {\doibase 10.1126/science.123884}
  {\bibfield  {journal} {\bibinfo  {journal} {Science}\ }\textbf {\bibinfo
  {volume} {342}},\ \bibinfo {pages} {212} (\bibinfo {year}
  {2013})}\BibitemShut {NoStop}%
\bibitem [{\citenamefont {Tallinen}\ \emph {et~al.}(2016)\citenamefont
  {Tallinen}, \citenamefont {Chung}, \citenamefont {Rousseau}, \citenamefont
  {Girard}, \citenamefont {Lef\`evre},\ and\ \citenamefont
  {Mahadevan}}]{tall16}%
  \BibitemOpen
  \bibfield  {author} {\bibinfo {author} {\bibfnamefont {T.}~\bibnamefont
  {Tallinen}}, \bibinfo {author} {\bibfnamefont {J.~Y.}\ \bibnamefont {Chung}},
  \bibinfo {author} {\bibfnamefont {F.}~\bibnamefont {Rousseau}}, \bibinfo
  {author} {\bibfnamefont {N.}~\bibnamefont {Girard}}, \bibinfo {author}
  {\bibfnamefont {J.}~\bibnamefont {Lef\`evre}}, \ and\ \bibinfo {author}
  {\bibfnamefont {L.}~\bibnamefont {Mahadevan}},\ }\href {\doibase
  10.1038/nphys3632} {\bibfield  {journal} {\bibinfo  {journal} {Nat. Phys.}\
  }\textbf {\bibinfo {volume} {12}},\ \bibinfo {pages} {588} (\bibinfo {year}
  {2016})}\BibitemShut {NoStop}%
\bibitem [{\citenamefont {Hallatschek}\ \emph {et~al.}(2023)\citenamefont
  {Hallatschek}, \citenamefont {Datta}, \citenamefont {Drescher}, \citenamefont
  {Dunkel}, \citenamefont {Elgeti}, \citenamefont {Waclaw},\ and\ \citenamefont
  {Wingreen}}]{hallatschekProliferatingActiveMatter2023}%
  \BibitemOpen
  \bibfield  {author} {\bibinfo {author} {\bibfnamefont {O.}~\bibnamefont
  {Hallatschek}}, \bibinfo {author} {\bibfnamefont {S.~S.}\ \bibnamefont
  {Datta}}, \bibinfo {author} {\bibfnamefont {K.}~\bibnamefont {Drescher}},
  \bibinfo {author} {\bibfnamefont {J.}~\bibnamefont {Dunkel}}, \bibinfo
  {author} {\bibfnamefont {J.}~\bibnamefont {Elgeti}}, \bibinfo {author}
  {\bibfnamefont {B.}~\bibnamefont {Waclaw}}, \ and\ \bibinfo {author}
  {\bibfnamefont {N.~S.}\ \bibnamefont {Wingreen}},\ }\href {\doibase
  10.1038/s42254-023-00593-0} {\bibfield  {journal} {\bibinfo  {journal} {Nat.
  Rev. Phys.}\ }\textbf {\bibinfo {volume} {5}},\ \bibinfo {pages} {407}
  (\bibinfo {year} {2023})}\BibitemShut {NoStop}%
\bibitem [{\citenamefont {Liedekerke}\ \emph {et~al.}(2020)\citenamefont
  {Liedekerke}, \citenamefont {Neitsch}, \citenamefont {Johann}, \citenamefont
  {Warmt}, \citenamefont {Gonz\'alez-Valverde}, \citenamefont {Hoehme},
  \citenamefont {Grosser}, \citenamefont {Kaes},\ and\ \citenamefont
  {Drasdo}}]{lied20}%
  \BibitemOpen
  \bibfield  {author} {\bibinfo {author} {\bibfnamefont {P.~V.}\ \bibnamefont
  {Liedekerke}}, \bibinfo {author} {\bibfnamefont {J.}~\bibnamefont {Neitsch}},
  \bibinfo {author} {\bibfnamefont {T.}~\bibnamefont {Johann}}, \bibinfo
  {author} {\bibfnamefont {E.}~\bibnamefont {Warmt}}, \bibinfo {author}
  {\bibfnamefont {I.}~\bibnamefont {Gonz\'alez-Valverde}}, \bibinfo {author}
  {\bibfnamefont {S.}~\bibnamefont {Hoehme}}, \bibinfo {author} {\bibfnamefont
  {S.}~\bibnamefont {Grosser}}, \bibinfo {author} {\bibfnamefont
  {J.}~\bibnamefont {Kaes}}, \ and\ \bibinfo {author} {\bibfnamefont
  {D.}~\bibnamefont {Drasdo}},\ }\href {\doibase 10.1007/s10237-019-01204-7}
  {\bibfield  {journal} {\bibinfo  {journal} {Biomech. Model. Mechanobiol.}\
  }\textbf {\bibinfo {volume} {19}},\ \bibinfo {pages} {189} (\bibinfo {year}
  {2020})}\BibitemShut {NoStop}%
\bibitem [{\citenamefont {Okuda}\ and\ \citenamefont {Hiraiwa}(2023)}]{okud23}%
  \BibitemOpen
  \bibfield  {author} {\bibinfo {author} {\bibfnamefont {S.}~\bibnamefont
  {Okuda}}\ and\ \bibinfo {author} {\bibfnamefont {T.}~\bibnamefont
  {Hiraiwa}},\ }\href {\doibase 10.1140/epje/s10189-023-00315-5} {\bibfield
  {journal} {\bibinfo  {journal} {Eur. Phys. J. E}\ }\textbf {\bibinfo {volume}
  {46}},\ \bibinfo {pages} {56} (\bibinfo {year} {2023})}\BibitemShut {NoStop}%
\bibitem [{\citenamefont {Cuvelier}\ \emph {et~al.}(2023)\citenamefont
  {Cuvelier}, \citenamefont {Vangheel}, \citenamefont {Thiels}, \citenamefont
  {Ramon}, \citenamefont {Jelier},\ and\ \citenamefont {Smeets}}]{cuve23}%
  \BibitemOpen
  \bibfield  {author} {\bibinfo {author} {\bibfnamefont {M.}~\bibnamefont
  {Cuvelier}}, \bibinfo {author} {\bibfnamefont {J.}~\bibnamefont {Vangheel}},
  \bibinfo {author} {\bibfnamefont {W.}~\bibnamefont {Thiels}}, \bibinfo
  {author} {\bibfnamefont {H.}~\bibnamefont {Ramon}}, \bibinfo {author}
  {\bibfnamefont {R.}~\bibnamefont {Jelier}}, \ and\ \bibinfo {author}
  {\bibfnamefont {B.}~\bibnamefont {Smeets}},\ }\href {\doibase
  10.1016/j.bpj.2023.04.017} {\bibfield  {journal} {\bibinfo  {journal}
  {Biophys. J.}\ }\textbf {\bibinfo {volume} {122}},\ \bibinfo {pages} {1858}
  (\bibinfo {year} {2023})}\BibitemShut {NoStop}%
\bibitem [{\citenamefont {Ichbiah}\ \emph {et~al.}(2023)\citenamefont
  {Ichbiah}, \citenamefont {Delbary}, \citenamefont {McDougall}, \citenamefont
  {Dumollard},\ and\ \citenamefont {Turlier}}]{ichb23}%
  \BibitemOpen
  \bibfield  {author} {\bibinfo {author} {\bibfnamefont {S.}~\bibnamefont
  {Ichbiah}}, \bibinfo {author} {\bibfnamefont {F.}~\bibnamefont {Delbary}},
  \bibinfo {author} {\bibfnamefont {A.}~\bibnamefont {McDougall}}, \bibinfo
  {author} {\bibfnamefont {R.}~\bibnamefont {Dumollard}}, \ and\ \bibinfo
  {author} {\bibfnamefont {H.}~\bibnamefont {Turlier}},\ }\href {\doibase
  10.1038/s41592-023-02084-7} {\bibfield  {journal} {\bibinfo  {journal} {Nat.
  Methods}\ }\textbf {\bibinfo {volume} {20}},\ \bibinfo {pages} {1989}
  (\bibinfo {year} {2023})}\BibitemShut {NoStop}%
\bibitem [{\citenamefont {Runser}\ \emph {et~al.}(2024)\citenamefont {Runser},
  \citenamefont {Vetter},\ and\ \citenamefont {Iber}}]{runs24}%
  \BibitemOpen
  \bibfield  {author} {\bibinfo {author} {\bibfnamefont {S.}~\bibnamefont
  {Runser}}, \bibinfo {author} {\bibfnamefont {R.}~\bibnamefont {Vetter}}, \
  and\ \bibinfo {author} {\bibfnamefont {D.}~\bibnamefont {Iber}},\ }\href
  {\doibase 10.1038/s43588-024-00620-9} {\bibfield  {journal} {\bibinfo
  {journal} {Nat. Comput. Sci.}\ }\textbf {\bibinfo {volume} {4}},\ \bibinfo
  {pages} {299} (\bibinfo {year} {2024})}\BibitemShut {NoStop}%
\bibitem [{\citenamefont {Fletcher}\ \emph {et~al.}(2014)\citenamefont
  {Fletcher}, \citenamefont {Osterfield}, \citenamefont {Baker},\ and\
  \citenamefont {Shvartsman}}]{flet14}%
  \BibitemOpen
  \bibfield  {author} {\bibinfo {author} {\bibfnamefont {A.~G.}\ \bibnamefont
  {Fletcher}}, \bibinfo {author} {\bibfnamefont {M.}~\bibnamefont
  {Osterfield}}, \bibinfo {author} {\bibfnamefont {R.~E.}\ \bibnamefont
  {Baker}}, \ and\ \bibinfo {author} {\bibfnamefont {S.~Y.}\ \bibnamefont
  {Shvartsman}},\ }\href {\doibase 10.1016/j.bpj.2013.11.4498} {\bibfield
  {journal} {\bibinfo  {journal} {Biophys. J.}\ }\textbf {\bibinfo {volume}
  {106}},\ \bibinfo {pages} {2291} (\bibinfo {year} {2014})}\BibitemShut
  {NoStop}%
\bibitem [{\citenamefont {Alt}\ \emph {et~al.}(2017)\citenamefont {Alt},
  \citenamefont {Ganguly},\ and\ \citenamefont {Salbreux}}]{alt17}%
  \BibitemOpen
  \bibfield  {author} {\bibinfo {author} {\bibfnamefont {S.}~\bibnamefont
  {Alt}}, \bibinfo {author} {\bibfnamefont {P.}~\bibnamefont {Ganguly}}, \ and\
  \bibinfo {author} {\bibfnamefont {G.}~\bibnamefont {Salbreux}},\ }\href
  {\doibase 10.1098/rstb.2015.0520} {\bibfield  {journal} {\bibinfo  {journal}
  {Phil. Trans. R. Soc. B}\ }\textbf {\bibinfo {volume} {372}},\ \bibinfo
  {pages} {20150520} (\bibinfo {year} {2017})}\BibitemShut {NoStop}%
\bibitem [{\citenamefont {Basan}\ \emph {et~al.}(2011)\citenamefont {Basan},
  \citenamefont {Prost}, \citenamefont {Joanny},\ and\ \citenamefont
  {Elgeti}}]{basan2011}%
  \BibitemOpen
  \bibfield  {author} {\bibinfo {author} {\bibfnamefont {M.}~\bibnamefont
  {Basan}}, \bibinfo {author} {\bibfnamefont {J.}~\bibnamefont {Prost}},
  \bibinfo {author} {\bibfnamefont {J.-F.}\ \bibnamefont {Joanny}}, \ and\
  \bibinfo {author} {\bibfnamefont {J.}~\bibnamefont {Elgeti}},\ }\href
  {\doibase 10.1088/1478-3975/8/2/026014} {\bibfield  {journal} {\bibinfo
  {journal} {Phys. Biol.}\ }\textbf {\bibinfo {volume} {8}},\ \bibinfo {pages}
  {026014} (\bibinfo {year} {2011})}\BibitemShut {NoStop}%
\bibitem [{\citenamefont {Murisic}\ \emph {et~al.}(2015)\citenamefont
  {Murisic}, \citenamefont {Hakim}, \citenamefont {Kevrekidis}, \citenamefont
  {Shvartsman},\ and\ \citenamefont {Audoly}}]{muri15}%
  \BibitemOpen
  \bibfield  {author} {\bibinfo {author} {\bibfnamefont {N.}~\bibnamefont
  {Murisic}}, \bibinfo {author} {\bibfnamefont {V.}~\bibnamefont {Hakim}},
  \bibinfo {author} {\bibfnamefont {I.~G.}\ \bibnamefont {Kevrekidis}},
  \bibinfo {author} {\bibfnamefont {S.~Y.}\ \bibnamefont {Shvartsman}}, \ and\
  \bibinfo {author} {\bibfnamefont {B.}~\bibnamefont {Audoly}},\ }\href
  {\doibase 10.1016/j.bpj.2015.05.019} {\bibfield  {journal} {\bibinfo
  {journal} {Biophys. J.}\ }\textbf {\bibinfo {volume} {109}},\ \bibinfo
  {pages} {154} (\bibinfo {year} {2015})}\BibitemShut {NoStop}%
\bibitem [{\citenamefont {Luciano}\ \emph {et~al.}(2021)\citenamefont
  {Luciano}, \citenamefont {Xue}, \citenamefont {Vos}, \citenamefont
  {Redondo-Morata}, \citenamefont {Surin}, \citenamefont {Lafont},
  \citenamefont {Hannezo},\ and\ \citenamefont {Gabriele}}]{luci21}%
  \BibitemOpen
  \bibfield  {author} {\bibinfo {author} {\bibfnamefont {M.}~\bibnamefont
  {Luciano}}, \bibinfo {author} {\bibfnamefont {S.-L.}\ \bibnamefont {Xue}},
  \bibinfo {author} {\bibfnamefont {W.~H.~D.}\ \bibnamefont {Vos}}, \bibinfo
  {author} {\bibfnamefont {L.}~\bibnamefont {Redondo-Morata}}, \bibinfo
  {author} {\bibfnamefont {M.}~\bibnamefont {Surin}}, \bibinfo {author}
  {\bibfnamefont {F.}~\bibnamefont {Lafont}}, \bibinfo {author} {\bibfnamefont
  {E.}~\bibnamefont {Hannezo}}, \ and\ \bibinfo {author} {\bibfnamefont
  {S.}~\bibnamefont {Gabriele}},\ }\href {\doibase 10.1038/s41567-021-01374-1}
  {\bibfield  {journal} {\bibinfo  {journal} {Nat. Phys.}\ }\textbf {\bibinfo
  {volume} {17}},\ \bibinfo {pages} {1382} (\bibinfo {year}
  {2021})}\BibitemShut {NoStop}%
\bibitem [{\citenamefont {Ogita}\ \emph {et~al.}(2022)\citenamefont {Ogita},
  \citenamefont {Kondo}, \citenamefont {Ikawa}, \citenamefont {Uemura},
  \citenamefont {Ishihara},\ and\ \citenamefont {Sugimura}}]{ogit22}%
  \BibitemOpen
  \bibfield  {author} {\bibinfo {author} {\bibfnamefont {G.}~\bibnamefont
  {Ogita}}, \bibinfo {author} {\bibfnamefont {T.}~\bibnamefont {Kondo}},
  \bibinfo {author} {\bibfnamefont {K.}~\bibnamefont {Ikawa}}, \bibinfo
  {author} {\bibfnamefont {T.}~\bibnamefont {Uemura}}, \bibinfo {author}
  {\bibfnamefont {S.}~\bibnamefont {Ishihara}}, \ and\ \bibinfo {author}
  {\bibfnamefont {K.}~\bibnamefont {Sugimura}},\ }\href {\doibase
  10.1371/journal.pcbi.1010209} {\bibfield  {journal} {\bibinfo  {journal}
  {PLoS Comput. Biol.}\ }\textbf {\bibinfo {volume} {18}},\ \bibinfo {pages}
  {e1010209} (\bibinfo {year} {2022})}\BibitemShut {NoStop}%
\bibitem [{\citenamefont {Tetley}\ \emph {et~al.}(2019)\citenamefont {Tetley},
  \citenamefont {Staddon}, \citenamefont {Heller}, \citenamefont {Hoppe},
  \citenamefont {Banerjee},\ and\ \citenamefont {Mao}}]{tetl19}%
  \BibitemOpen
  \bibfield  {author} {\bibinfo {author} {\bibfnamefont {R.~J.}\ \bibnamefont
  {Tetley}}, \bibinfo {author} {\bibfnamefont {M.~F.}\ \bibnamefont {Staddon}},
  \bibinfo {author} {\bibfnamefont {D.}~\bibnamefont {Heller}}, \bibinfo
  {author} {\bibfnamefont {A.}~\bibnamefont {Hoppe}}, \bibinfo {author}
  {\bibfnamefont {S.}~\bibnamefont {Banerjee}}, \ and\ \bibinfo {author}
  {\bibfnamefont {Y.}~\bibnamefont {Mao}},\ }\href {\doibase
  /10.1038/s41567-019-0618-1} {\bibfield  {journal} {\bibinfo  {journal} {Nat.
  Phys.}\ }\textbf {\bibinfo {volume} {15}},\ \bibinfo {pages} {1195} (\bibinfo
  {year} {2019})}\BibitemShut {NoStop}%
\bibitem [{\citenamefont {Prakash}\ \emph {et~al.}(2021)\citenamefont
  {Prakash}, \citenamefont {Bull},\ and\ \citenamefont {Prakash}}]{prak21}%
  \BibitemOpen
  \bibfield  {author} {\bibinfo {author} {\bibfnamefont {V.~N.}\ \bibnamefont
  {Prakash}}, \bibinfo {author} {\bibfnamefont {M.~S.}\ \bibnamefont {Bull}}, \
  and\ \bibinfo {author} {\bibfnamefont {M.}~\bibnamefont {Prakash}},\ }\href
  {\doibase 10.1038/s41567-020-01134-7} {\bibfield  {journal} {\bibinfo
  {journal} {Nat. Phys.}\ }\textbf {\bibinfo {volume} {17}},\ \bibinfo {pages}
  {504} (\bibinfo {year} {2021})}\BibitemShut {NoStop}%
\bibitem [{\citenamefont {Xu}\ \emph {et~al.}(2022)\citenamefont {Xu},
  \citenamefont {Xu}, \citenamefont {Li}, \citenamefont {He}, \citenamefont
  {Li},\ and\ \citenamefont {Ji}}]{xu22}%
  \BibitemOpen
  \bibfield  {author} {\bibinfo {author} {\bibfnamefont {J.}~\bibnamefont
  {Xu}}, \bibinfo {author} {\bibfnamefont {X.}~\bibnamefont {Xu}}, \bibinfo
  {author} {\bibfnamefont {X.}~\bibnamefont {Li}}, \bibinfo {author}
  {\bibfnamefont {S.}~\bibnamefont {He}}, \bibinfo {author} {\bibfnamefont
  {D.}~\bibnamefont {Li}}, \ and\ \bibinfo {author} {\bibfnamefont
  {B.}~\bibnamefont {Ji}},\ }\href {\doibase 10.1016/j.bpj.2021.12.015}
  {\bibfield  {journal} {\bibinfo  {journal} {Biophys. J.}\ }\textbf {\bibinfo
  {volume} {121}},\ \bibinfo {pages} {288} (\bibinfo {year}
  {2022})}\BibitemShut {NoStop}%
\bibitem [{\citenamefont {Sonam}\ \emph {et~al.}(2023)\citenamefont {Sonam},
  \citenamefont {Balasubramaniam}, \citenamefont {Lin}, \citenamefont {Ivan},
  \citenamefont {Pi-Jaum\'a}, \citenamefont {Jebane}, \citenamefont {Karnat},
  \citenamefont {Toyama}, \citenamefont {Marcq}, \citenamefont {Prost},
  \citenamefont {M\'ege}, \citenamefont {Rupprecht},\ and\ \citenamefont
  {Ladoux}}]{sona23}%
  \BibitemOpen
  \bibfield  {author} {\bibinfo {author} {\bibfnamefont {S.}~\bibnamefont
  {Sonam}}, \bibinfo {author} {\bibfnamefont {L.}~\bibnamefont
  {Balasubramaniam}}, \bibinfo {author} {\bibfnamefont {S.-Z.}\ \bibnamefont
  {Lin}}, \bibinfo {author} {\bibfnamefont {Y.~M.~Y.}\ \bibnamefont {Ivan}},
  \bibinfo {author} {\bibfnamefont {I.}~\bibnamefont {Pi-Jaum\'a}}, \bibinfo
  {author} {\bibfnamefont {C.}~\bibnamefont {Jebane}}, \bibinfo {author}
  {\bibfnamefont {M.}~\bibnamefont {Karnat}}, \bibinfo {author} {\bibfnamefont
  {Y.}~\bibnamefont {Toyama}}, \bibinfo {author} {\bibfnamefont
  {P.}~\bibnamefont {Marcq}}, \bibinfo {author} {\bibfnamefont
  {J.}~\bibnamefont {Prost}}, \bibinfo {author} {\bibfnamefont {R.-M.}\
  \bibnamefont {M\'ege}}, \bibinfo {author} {\bibfnamefont {J.-F.}\
  \bibnamefont {Rupprecht}}, \ and\ \bibinfo {author} {\bibfnamefont
  {B.}~\bibnamefont {Ladoux}},\ }\href {\doibase 10.1038/s41567-022-01826-2}
  {\bibfield  {journal} {\bibinfo  {journal} {Nat. Phys.}\ }\textbf {\bibinfo
  {volume} {19}},\ \bibinfo {pages} {132} (\bibinfo {year} {2023})}\BibitemShut
  {NoStop}%
\bibitem [{\citenamefont {Lv}\ \emph {et~al.}(2024)\citenamefont {Lv},
  \citenamefont {Chen}, \citenamefont {Chen}, \citenamefont {Liu},
  \citenamefont {Wang}, \citenamefont {Bai}, \citenamefont {Lv}, \citenamefont
  {Li}, \citenamefont {Shao}, \citenamefont {Feng},\ and\ \citenamefont
  {Li}}]{lv24}%
  \BibitemOpen
  \bibfield  {author} {\bibinfo {author} {\bibfnamefont {J.-Q.}\ \bibnamefont
  {Lv}}, \bibinfo {author} {\bibfnamefont {P.-C.}\ \bibnamefont {Chen}},
  \bibinfo {author} {\bibfnamefont {Y.-P.}\ \bibnamefont {Chen}}, \bibinfo
  {author} {\bibfnamefont {H.-Y.}\ \bibnamefont {Liu}}, \bibinfo {author}
  {\bibfnamefont {S.-D.}\ \bibnamefont {Wang}}, \bibinfo {author}
  {\bibfnamefont {J.}~\bibnamefont {Bai}}, \bibinfo {author} {\bibfnamefont
  {C.-L.}\ \bibnamefont {Lv}}, \bibinfo {author} {\bibfnamefont
  {Y.}~\bibnamefont {Li}}, \bibinfo {author} {\bibfnamefont {Y.}~\bibnamefont
  {Shao}}, \bibinfo {author} {\bibfnamefont {X.-Q.}\ \bibnamefont {Feng}}, \
  and\ \bibinfo {author} {\bibfnamefont {B.}~\bibnamefont {Li}},\ }\href
  {\doibase 10.1038/s41567-024-02504-1} {\bibfield  {journal} {\bibinfo
  {journal} {Nat. Phys.}\ }\textbf {\bibinfo {volume} {20}},\ \bibinfo {pages}
  {1313} (\bibinfo {year} {2024})}\BibitemShut {NoStop}%
\bibitem [{\citenamefont {Guerrero}\ and\ \citenamefont
  {Perez-Carrasco}(2023)}]{guer23}%
  \BibitemOpen
  \bibfield  {author} {\bibinfo {author} {\bibfnamefont {P.}~\bibnamefont
  {Guerrero}}\ and\ \bibinfo {author} {\bibfnamefont {R.}~\bibnamefont
  {Perez-Carrasco}},\ }\href {\doibase 10.1098/rstb.2023.0051} {\bibfield
  {journal} {\bibinfo  {journal} {Phil. Trans. R. Soc. B}\ }\textbf {\bibinfo
  {volume} {379}},\ \bibinfo {pages} {20230051} (\bibinfo {year}
  {2023})}\BibitemShut {NoStop}%
\bibitem [{\citenamefont {Hirashima}\ and\ \citenamefont
  {Matsuda}(2024)}]{hira24}%
  \BibitemOpen
  \bibfield  {author} {\bibinfo {author} {\bibfnamefont {T.}~\bibnamefont
  {Hirashima}}\ and\ \bibinfo {author} {\bibfnamefont {M.}~\bibnamefont
  {Matsuda}},\ }\href {\doibase 10.1016/j.cub.2023.12.049} {\bibfield
  {journal} {\bibinfo  {journal} {Curr. Biol.}\ }\textbf {\bibinfo {volume}
  {34}},\ \bibinfo {pages} {683} (\bibinfo {year} {2024})}\BibitemShut
  {NoStop}%
\bibitem [{\citenamefont {Honda}\ \emph {et~al.}(2004)\citenamefont {Honda},
  \citenamefont {Tanemura},\ and\ \citenamefont {Nagai}}]{hond04}%
  \BibitemOpen
  \bibfield  {author} {\bibinfo {author} {\bibfnamefont {H.}~\bibnamefont
  {Honda}}, \bibinfo {author} {\bibfnamefont {M.}~\bibnamefont {Tanemura}}, \
  and\ \bibinfo {author} {\bibfnamefont {T.}~\bibnamefont {Nagai}},\ }\href
  {\doibase 10.1016/j.jtbi.2003.10.001} {\bibfield  {journal} {\bibinfo
  {journal} {J. Theor. Biol.}\ }\textbf {\bibinfo {volume} {226}},\ \bibinfo
  {pages} {439} (\bibinfo {year} {2004})}\BibitemShut {NoStop}%
\bibitem [{\citenamefont {Okuda}\ \emph {et~al.}(2013)\citenamefont {Okuda},
  \citenamefont {Inoue}, \citenamefont {Eiraku}, \citenamefont {Sasai},\ and\
  \citenamefont {Adachi}}]{okud13}%
  \BibitemOpen
  \bibfield  {author} {\bibinfo {author} {\bibfnamefont {S.}~\bibnamefont
  {Okuda}}, \bibinfo {author} {\bibfnamefont {Y.}~\bibnamefont {Inoue}},
  \bibinfo {author} {\bibfnamefont {M.}~\bibnamefont {Eiraku}}, \bibinfo
  {author} {\bibfnamefont {Y.}~\bibnamefont {Sasai}}, \ and\ \bibinfo {author}
  {\bibfnamefont {T.}~\bibnamefont {Adachi}},\ }\href {\doibase
  10.1007/s10237-012-0430-7} {\bibfield  {journal} {\bibinfo  {journal}
  {Biomech. Model. Mechanobiol.}\ }\textbf {\bibinfo {volume} {12}},\ \bibinfo
  {pages} {627} (\bibinfo {year} {2013})}\BibitemShut {NoStop}%
\bibitem [{\citenamefont {Krajnc}\ \emph {et~al.}(2018)\citenamefont {Krajnc},
  \citenamefont {Dasgupta}, \citenamefont {Ziherl},\ and\ \citenamefont
  {Prost}}]{kraj18}%
  \BibitemOpen
  \bibfield  {author} {\bibinfo {author} {\bibfnamefont {M.}~\bibnamefont
  {Krajnc}}, \bibinfo {author} {\bibfnamefont {S.}~\bibnamefont {Dasgupta}},
  \bibinfo {author} {\bibfnamefont {P.}~\bibnamefont {Ziherl}}, \ and\ \bibinfo
  {author} {\bibfnamefont {J.}~\bibnamefont {Prost}},\ }\href {\doibase
  10.1103/PhysRevE.98.022409} {\bibfield  {journal} {\bibinfo  {journal} {Phys.
  Rev. E}\ }\textbf {\bibinfo {volume} {98}},\ \bibinfo {pages} {022409}
  (\bibinfo {year} {2018})}\BibitemShut {NoStop}%
\bibitem [{\citenamefont {Okuda}\ \emph {et~al.}(2018)\citenamefont {Okuda},
  \citenamefont {Miura}, \citenamefont {Inoue}, \citenamefont {Adachi},\ and\
  \citenamefont {Eiraku}}]{okud18}%
  \BibitemOpen
  \bibfield  {author} {\bibinfo {author} {\bibfnamefont {S.}~\bibnamefont
  {Okuda}}, \bibinfo {author} {\bibfnamefont {T.}~\bibnamefont {Miura}},
  \bibinfo {author} {\bibfnamefont {Y.}~\bibnamefont {Inoue}}, \bibinfo
  {author} {\bibfnamefont {T.}~\bibnamefont {Adachi}}, \ and\ \bibinfo {author}
  {\bibfnamefont {M.}~\bibnamefont {Eiraku}},\ }\href {\doibase
  10.1038/s41598-018-20678-6} {\bibfield  {journal} {\bibinfo  {journal} {Sci.
  Rep.}\ }\textbf {\bibinfo {volume} {8}},\ \bibinfo {pages} {2386} (\bibinfo
  {year} {2018})}\BibitemShut {NoStop}%
\bibitem [{\citenamefont {Inoue}\ \emph {et~al.}(2020)\citenamefont {Inoue},
  \citenamefont {Tateo},\ and\ \citenamefont {Adachi}}]{inou20}%
  \BibitemOpen
  \bibfield  {author} {\bibinfo {author} {\bibfnamefont {Y.}~\bibnamefont
  {Inoue}}, \bibinfo {author} {\bibfnamefont {I.}~\bibnamefont {Tateo}}, \ and\
  \bibinfo {author} {\bibfnamefont {T.}~\bibnamefont {Adachi}},\ }\href
  {\doibase 10.1007/s10237-019-01249-8} {\bibfield  {journal} {\bibinfo
  {journal} {Biomech. Model. Mechanobiol.}\ }\textbf {\bibinfo {volume} {19}},\
  \bibinfo {pages} {815} (\bibinfo {year} {2020})}\BibitemShut {NoStop}%
\bibitem [{\citenamefont {Zhang}\ and\ \citenamefont {Schwarz}(2022)}]{zhan22}%
  \BibitemOpen
  \bibfield  {author} {\bibinfo {author} {\bibfnamefont {T.}~\bibnamefont
  {Zhang}}\ and\ \bibinfo {author} {\bibfnamefont {J.~M.}\ \bibnamefont
  {Schwarz}},\ }\href {\doibase 10.1103/PhysRevResearch.4.043148} {\bibfield
  {journal} {\bibinfo  {journal} {Phys. Rev. Res.}\ }\textbf {\bibinfo {volume}
  {4}},\ \bibinfo {pages} {043148} (\bibinfo {year} {2022})}\BibitemShut
  {NoStop}%
\bibitem [{\citenamefont {Meineke}\ \emph {et~al.}(2001)\citenamefont
  {Meineke}, \citenamefont {Potten},\ and\ \citenamefont {Loeffler}}]{mein01}%
  \BibitemOpen
  \bibfield  {author} {\bibinfo {author} {\bibfnamefont {F.~A.}\ \bibnamefont
  {Meineke}}, \bibinfo {author} {\bibfnamefont {C.~S.}\ \bibnamefont {Potten}},
  \ and\ \bibinfo {author} {\bibfnamefont {M.}~\bibnamefont {Loeffler}},\
  }\href {\doibase 10.1046/j.0960-7722.2001.00216.x} {\bibfield  {journal}
  {\bibinfo  {journal} {Cell Prolif.}\ }\textbf {\bibinfo {volume} {34}},\
  \bibinfo {pages} {253} (\bibinfo {year} {2001})}\BibitemShut {NoStop}%
\bibitem [{\citenamefont {Bi}\ \emph {et~al.}(2016)\citenamefont {Bi},
  \citenamefont {Yang}, \citenamefont {Marchetti},\ and\ \citenamefont
  {Manning}}]{dape16}%
  \BibitemOpen
  \bibfield  {author} {\bibinfo {author} {\bibfnamefont {D.}~\bibnamefont
  {Bi}}, \bibinfo {author} {\bibfnamefont {X.}~\bibnamefont {Yang}}, \bibinfo
  {author} {\bibfnamefont {M.~C.}\ \bibnamefont {Marchetti}}, \ and\ \bibinfo
  {author} {\bibfnamefont {M.~L.}\ \bibnamefont {Manning}},\ }\href {\doibase
  10.1103/PhysRevX.6.021011} {\bibfield  {journal} {\bibinfo  {journal} {Phys.
  Rev. X}\ }\textbf {\bibinfo {volume} {6}},\ \bibinfo {pages} {021011}
  (\bibinfo {year} {2016})}\BibitemShut {NoStop}%
\bibitem [{\citenamefont {Mimura}\ and\ \citenamefont {Inoue}(2023)}]{mimu23}%
  \BibitemOpen
  \bibfield  {author} {\bibinfo {author} {\bibfnamefont {T.}~\bibnamefont
  {Mimura}}\ and\ \bibinfo {author} {\bibfnamefont {Y.}~\bibnamefont {Inoue}},\
  }\href {\doibase 10.1016/j.jtbi.2023.111560} {\bibfield  {journal} {\bibinfo
  {journal} {J. Theor. Biol.}\ }\textbf {\bibinfo {volume} {571}},\ \bibinfo
  {pages} {111560} (\bibinfo {year} {2023})}\BibitemShut {NoStop}%
\bibitem [{\citenamefont {Ranft}\ \emph {et~al.}(2010)\citenamefont {Ranft},
  \citenamefont {Basan}, \citenamefont {Elgeti}, \citenamefont {Joanny},
  \citenamefont {Prost},\ and\ \citenamefont
  {Julicher}}]{ranftFluidizationTissuesCell2010}%
  \BibitemOpen
  \bibfield  {author} {\bibinfo {author} {\bibfnamefont {J.}~\bibnamefont
  {Ranft}}, \bibinfo {author} {\bibfnamefont {M.}~\bibnamefont {Basan}},
  \bibinfo {author} {\bibfnamefont {J.}~\bibnamefont {Elgeti}}, \bibinfo
  {author} {\bibfnamefont {J.-F.}\ \bibnamefont {Joanny}}, \bibinfo {author}
  {\bibfnamefont {J.}~\bibnamefont {Prost}}, \ and\ \bibinfo {author}
  {\bibfnamefont {F.}~\bibnamefont {Julicher}},\ }\href {\doibase
  10.1073/pnas.1011086107} {\bibfield  {journal} {\bibinfo  {journal} {Proc.\
  Natl.\ Acad.\ Sci.\ USA}\ }\textbf {\bibinfo {volume} {107}},\ \bibinfo
  {pages} {20863} (\bibinfo {year} {2010})}\BibitemShut {NoStop}%
\bibitem [{\citenamefont {Montel}\ \emph {et~al.}(2011)\citenamefont {Montel},
  \citenamefont {Delarue}, \citenamefont {Elgeti}, \citenamefont {Malaquin},
  \citenamefont {Basan}, \citenamefont {Risler}, \citenamefont {Cabane},
  \citenamefont {Vignjevic}, \citenamefont {Prost}, \citenamefont {Cappello},\
  and\ \citenamefont {Joanny}}]{montelStressClampExperiments2011}%
  \BibitemOpen
  \bibfield  {author} {\bibinfo {author} {\bibfnamefont {F.}~\bibnamefont
  {Montel}}, \bibinfo {author} {\bibfnamefont {M.}~\bibnamefont {Delarue}},
  \bibinfo {author} {\bibfnamefont {J.}~\bibnamefont {Elgeti}}, \bibinfo
  {author} {\bibfnamefont {L.}~\bibnamefont {Malaquin}}, \bibinfo {author}
  {\bibfnamefont {M.}~\bibnamefont {Basan}}, \bibinfo {author} {\bibfnamefont
  {T.}~\bibnamefont {Risler}}, \bibinfo {author} {\bibfnamefont
  {B.}~\bibnamefont {Cabane}}, \bibinfo {author} {\bibfnamefont
  {D.}~\bibnamefont {Vignjevic}}, \bibinfo {author} {\bibfnamefont
  {J.}~\bibnamefont {Prost}}, \bibinfo {author} {\bibfnamefont
  {G.}~\bibnamefont {Cappello}}, \ and\ \bibinfo {author} {\bibfnamefont
  {J.-F.}\ \bibnamefont {Joanny}},\ }\href {\doibase
  10.1103/PhysRevLett.107.188102} {\bibfield  {journal} {\bibinfo  {journal}
  {Phys. Rev. Lett.}\ }\textbf {\bibinfo {volume} {107}},\ \bibinfo {pages}
  {188102} (\bibinfo {year} {2011})}\BibitemShut {NoStop}%
\bibitem [{\citenamefont {Montel}\ \emph {et~al.}(2012)\citenamefont {Montel},
  \citenamefont {Delarue}, \citenamefont {Elgeti}, \citenamefont {Vignjevic},
  \citenamefont {Cappello},\ and\ \citenamefont
  {Prost}}]{montelIsotropicStressReduces2012}%
  \BibitemOpen
  \bibfield  {author} {\bibinfo {author} {\bibfnamefont {F.}~\bibnamefont
  {Montel}}, \bibinfo {author} {\bibfnamefont {M.}~\bibnamefont {Delarue}},
  \bibinfo {author} {\bibfnamefont {J.}~\bibnamefont {Elgeti}}, \bibinfo
  {author} {\bibfnamefont {D.}~\bibnamefont {Vignjevic}}, \bibinfo {author}
  {\bibfnamefont {G.}~\bibnamefont {Cappello}}, \ and\ \bibinfo {author}
  {\bibfnamefont {J.}~\bibnamefont {Prost}},\ }\href {\doibase
  10.1088/1367-2630/14/5/055008} {\bibfield  {journal} {\bibinfo  {journal}
  {New J. Phys.}\ }\textbf {\bibinfo {volume} {14}},\ \bibinfo {pages} {055008}
  (\bibinfo {year} {2012})}\BibitemShut {NoStop}%
\bibitem [{\citenamefont {Delarue}\ \emph {et~al.}(2013)\citenamefont
  {Delarue}, \citenamefont {Montel}, \citenamefont {Caen}, \citenamefont
  {Elgeti}, \citenamefont {Siaugue}, \citenamefont {Vignjevic}, \citenamefont
  {Prost}, \citenamefont {Joanny},\ and\ \citenamefont
  {Cappello}}]{delarueMechanicalControlCell2013}%
  \BibitemOpen
  \bibfield  {author} {\bibinfo {author} {\bibfnamefont {M.}~\bibnamefont
  {Delarue}}, \bibinfo {author} {\bibfnamefont {F.}~\bibnamefont {Montel}},
  \bibinfo {author} {\bibfnamefont {O.}~\bibnamefont {Caen}}, \bibinfo {author}
  {\bibfnamefont {J.}~\bibnamefont {Elgeti}}, \bibinfo {author} {\bibfnamefont
  {J.-M.}\ \bibnamefont {Siaugue}}, \bibinfo {author} {\bibfnamefont
  {D.}~\bibnamefont {Vignjevic}}, \bibinfo {author} {\bibfnamefont
  {J.}~\bibnamefont {Prost}}, \bibinfo {author} {\bibfnamefont {J.-F.}\
  \bibnamefont {Joanny}}, \ and\ \bibinfo {author} {\bibfnamefont
  {G.}~\bibnamefont {Cappello}},\ }\href {\doibase
  10.1103/PhysRevLett.110.138103} {\bibfield  {journal} {\bibinfo  {journal}
  {Phys. Rev. Lett.}\ }\textbf {\bibinfo {volume} {110}},\ \bibinfo {pages}
  {138103} (\bibinfo {year} {2013})}\BibitemShut {NoStop}%
\bibitem [{\citenamefont {B\"uscher}\ \emph
  {et~al.}(2020{\natexlab{a}})\citenamefont {B\"uscher}, \citenamefont {Ganai},
  \citenamefont {Gompper},\ and\ \citenamefont
  {Elgeti}}]{buscherTissueEvolutionMechanical2020}%
  \BibitemOpen
  \bibfield  {author} {\bibinfo {author} {\bibfnamefont {T.}~\bibnamefont
  {B\"uscher}}, \bibinfo {author} {\bibfnamefont {N.}~\bibnamefont {Ganai}},
  \bibinfo {author} {\bibfnamefont {G.}~\bibnamefont {Gompper}}, \ and\
  \bibinfo {author} {\bibfnamefont {J.}~\bibnamefont {Elgeti}},\ }\href
  {\doibase 10.1088/1367-2630/ab74a5} {\bibfield  {journal} {\bibinfo
  {journal} {New J. Phys.}\ }\textbf {\bibinfo {volume} {22}},\ \bibinfo
  {pages} {033048} (\bibinfo {year} {2020}{\natexlab{a}})}\BibitemShut
  {NoStop}%
\bibitem [{\citenamefont {B\"uscher}\ \emph
  {et~al.}(2020{\natexlab{b}})\citenamefont {B\"uscher}, \citenamefont {Diez},
  \citenamefont {Gompper},\ and\ \citenamefont
  {Elgeti}}]{buscherInstabilityFingeringInterfaces2020}%
  \BibitemOpen
  \bibfield  {author} {\bibinfo {author} {\bibfnamefont {T.}~\bibnamefont
  {B\"uscher}}, \bibinfo {author} {\bibfnamefont {A.~L.}\ \bibnamefont {Diez}},
  \bibinfo {author} {\bibfnamefont {G.}~\bibnamefont {Gompper}}, \ and\
  \bibinfo {author} {\bibfnamefont {J.}~\bibnamefont {Elgeti}},\ }\href
  {\doibase 10.1088/1367-2630/ab9e88} {\bibfield  {journal} {\bibinfo
  {journal} {New J. Phys.}\ }\textbf {\bibinfo {volume} {22}},\ \bibinfo
  {pages} {083005} (\bibinfo {year} {2020}{\natexlab{b}})}\BibitemShut
  {NoStop}%
\bibitem [{\citenamefont {Ganai}\ \emph {et~al.}(2019)\citenamefont {Ganai},
  \citenamefont {B\"uscher}, \citenamefont {Gompper},\ and\ \citenamefont
  {Elgeti}}]{ganaiMechanicsTissueCompetition2019}%
  \BibitemOpen
  \bibfield  {author} {\bibinfo {author} {\bibfnamefont {N.}~\bibnamefont
  {Ganai}}, \bibinfo {author} {\bibfnamefont {T.}~\bibnamefont {B\"uscher}},
  \bibinfo {author} {\bibfnamefont {G.}~\bibnamefont {Gompper}}, \ and\
  \bibinfo {author} {\bibfnamefont {J.}~\bibnamefont {Elgeti}},\ }\href
  {\doibase 10.1088/1367-2630/ab2475} {\bibfield  {journal} {\bibinfo
  {journal} {New J. Phys.}\ }\textbf {\bibinfo {volume} {21}},\ \bibinfo
  {pages} {063017} (\bibinfo {year} {2019})}\BibitemShut {NoStop}%
\bibitem [{\citenamefont {Podewitz}\ \emph {et~al.}(2016)\citenamefont
  {Podewitz}, \citenamefont {J\"ulicher}, \citenamefont {Gompper},\ and\
  \citenamefont {Elgeti}}]{podewitzInterfaceDynamicsCompeting2016}%
  \BibitemOpen
  \bibfield  {author} {\bibinfo {author} {\bibfnamefont {N.}~\bibnamefont
  {Podewitz}}, \bibinfo {author} {\bibfnamefont {F.}~\bibnamefont
  {J\"ulicher}}, \bibinfo {author} {\bibfnamefont {G.}~\bibnamefont {Gompper}},
  \ and\ \bibinfo {author} {\bibfnamefont {J.}~\bibnamefont {Elgeti}},\ }\href
  {\doibase 10.1088/1367-2630/18/8/083020} {\bibfield  {journal} {\bibinfo
  {journal} {New J. Phys.}\ }\textbf {\bibinfo {volume} {18}},\ \bibinfo
  {pages} {083020} (\bibinfo {year} {2016})}\BibitemShut {NoStop}%
\bibitem [{\citenamefont {Basan}\ \emph {et~al.}(2013)\citenamefont {Basan},
  \citenamefont {Elgeti}, \citenamefont {Hannezo}, \citenamefont {Rappel},\
  and\ \citenamefont {Levine}}]{basanAlignmentCellularMotility2013}%
  \BibitemOpen
  \bibfield  {author} {\bibinfo {author} {\bibfnamefont {M.}~\bibnamefont
  {Basan}}, \bibinfo {author} {\bibfnamefont {J.}~\bibnamefont {Elgeti}},
  \bibinfo {author} {\bibfnamefont {E.}~\bibnamefont {Hannezo}}, \bibinfo
  {author} {\bibfnamefont {W.-J.}\ \bibnamefont {Rappel}}, \ and\ \bibinfo
  {author} {\bibfnamefont {H.}~\bibnamefont {Levine}},\ }\href {\doibase
  10.1073/pnas.1219937110} {\bibfield  {journal} {\bibinfo  {journal} {Proc.\
  Natl.\ Acad.\ Sci.\ USA}\ }\textbf {\bibinfo {volume} {110}},\ \bibinfo
  {pages} {2452} (\bibinfo {year} {2013})}\BibitemShut {NoStop}%
\bibitem [{\citenamefont {Marel}\ \emph {et~al.}(2014)\citenamefont {Marel},
  \citenamefont {Podewitz}, \citenamefont {Zorn}, \citenamefont {R\"adler},\
  and\ \citenamefont {Elgeti}}]{marelAlignmentCellDivision2014}%
  \BibitemOpen
  \bibfield  {author} {\bibinfo {author} {\bibfnamefont {A.-K.}\ \bibnamefont
  {Marel}}, \bibinfo {author} {\bibfnamefont {N.}~\bibnamefont {Podewitz}},
  \bibinfo {author} {\bibfnamefont {M.}~\bibnamefont {Zorn}}, \bibinfo {author}
  {\bibfnamefont {J.~O.}\ \bibnamefont {R\"adler}}, \ and\ \bibinfo {author}
  {\bibfnamefont {J.}~\bibnamefont {Elgeti}},\ }\href {\doibase
  10.1088/1367-2630/16/11/115005} {\bibfield  {journal} {\bibinfo  {journal}
  {New J. Phys.}\ }\textbf {\bibinfo {volume} {16}},\ \bibinfo {pages} {115005}
  (\bibinfo {year} {2014})}\BibitemShut {NoStop}%
\bibitem [{\citenamefont {Kr\"amer}\ \emph {et~al.}(2024)\citenamefont
  {Kr\"amer}, \citenamefont {Hannezo}, \citenamefont {Gompper},\ and\
  \citenamefont {Elgeti}}]{kramerMechanicallydrivenStemCell2024}%
  \BibitemOpen
  \bibfield  {author} {\bibinfo {author} {\bibfnamefont {J.~C.}\ \bibnamefont
  {Kr\"amer}}, \bibinfo {author} {\bibfnamefont {E.}~\bibnamefont {Hannezo}},
  \bibinfo {author} {\bibfnamefont {G.}~\bibnamefont {Gompper}}, \ and\
  \bibinfo {author} {\bibfnamefont {J.}~\bibnamefont {Elgeti}},\ }\href
  {\doibase 10.21468/SciPostPhys.16.4.097} {\bibfield  {journal} {\bibinfo
  {journal} {SciPost Phys.}\ }\textbf {\bibinfo {volume} {16}},\ \bibinfo
  {pages} {097} (\bibinfo {year} {2024})}\BibitemShut {NoStop}%
\bibitem [{\citenamefont {Hornung}\ \emph {et~al.}(2018)\citenamefont
  {Hornung}, \citenamefont {Gr\"unberger}, \citenamefont {Westerwalbesloh},
  \citenamefont {Kohlheyer}, \citenamefont {Gompper},\ and\ \citenamefont
  {Elgeti}}]{hornungQuantitativeModellingNutrientlimited2018}%
  \BibitemOpen
  \bibfield  {author} {\bibinfo {author} {\bibfnamefont {R.}~\bibnamefont
  {Hornung}}, \bibinfo {author} {\bibfnamefont {A.}~\bibnamefont
  {Gr\"unberger}}, \bibinfo {author} {\bibfnamefont {C.}~\bibnamefont
  {Westerwalbesloh}}, \bibinfo {author} {\bibfnamefont {D.}~\bibnamefont
  {Kohlheyer}}, \bibinfo {author} {\bibfnamefont {G.}~\bibnamefont {Gompper}},
  \ and\ \bibinfo {author} {\bibfnamefont {J.}~\bibnamefont {Elgeti}},\ }\href
  {\doibase 10.1098/rsif.2017.0713} {\bibfield  {journal} {\bibinfo  {journal}
  {J. R. Soc. Interface}\ }\textbf {\bibinfo {volume} {15}},\ \bibinfo {pages}
  {20170713} (\bibinfo {year} {2018})}\BibitemShut {NoStop}%
\bibitem [{\citenamefont {Noguchi}\ and\ \citenamefont
  {Gompper}(2006{\natexlab{a}})}]{nogu06}%
  \BibitemOpen
  \bibfield  {author} {\bibinfo {author} {\bibfnamefont {H.}~\bibnamefont
  {Noguchi}}\ and\ \bibinfo {author} {\bibfnamefont {G.}~\bibnamefont
  {Gompper}},\ }\href {\doibase 10.1103/PhysRevE.73.021903} {\bibfield
  {journal} {\bibinfo  {journal} {Phys.\ Rev.\ E}\ }\textbf {\bibinfo {volume}
  {73}},\ \bibinfo {pages} {021903} (\bibinfo {year}
  {2006}{\natexlab{a}})}\BibitemShut {NoStop}%
\bibitem [{\citenamefont {Noguchi}(2010)}]{nogu10}%
  \BibitemOpen
  \bibfield  {author} {\bibinfo {author} {\bibfnamefont {H.}~\bibnamefont
  {Noguchi}},\ }\href {\doibase 10.1143/JPSJ.79.024801} {\bibfield  {journal}
  {\bibinfo  {journal} {J.\ Phys.\ Soc.\ Jpn.}\ }\textbf {\bibinfo {volume}
  {79}},\ \bibinfo {pages} {024801} (\bibinfo {year} {2010})}\BibitemShut
  {NoStop}%
\bibitem [{\citenamefont {Drouffe}\ \emph {et~al.}(1991)\citenamefont
  {Drouffe}, \citenamefont {Maggs},\ and\ \citenamefont {Leibler}}]{drou91}%
  \BibitemOpen
  \bibfield  {author} {\bibinfo {author} {\bibfnamefont {J.~M.}\ \bibnamefont
  {Drouffe}}, \bibinfo {author} {\bibfnamefont {A.~C.}\ \bibnamefont {Maggs}},
  \ and\ \bibinfo {author} {\bibfnamefont {S.}~\bibnamefont {Leibler}},\
  }\href@noop {} {\bibfield  {journal} {\bibinfo  {journal} {Science}\ }\textbf
  {\bibinfo {volume} {254}},\ \bibinfo {pages} {1353} (\bibinfo {year}
  {1991})}\BibitemShut {NoStop}%
\bibitem [{\citenamefont {Noguchi}\ and\ \citenamefont
  {Gompper}(2006{\natexlab{b}})}]{nogu06a}%
  \BibitemOpen
  \bibfield  {author} {\bibinfo {author} {\bibfnamefont {H.}~\bibnamefont
  {Noguchi}}\ and\ \bibinfo {author} {\bibfnamefont {G.}~\bibnamefont
  {Gompper}},\ }\href {\doibase 10.1063/1.2358983} {\bibfield  {journal}
  {\bibinfo  {journal} {J.\ Chem.\ Phys.}\ }\textbf {\bibinfo {volume} {125}},\
  \bibinfo {pages} {164908} (\bibinfo {year} {2006}{\natexlab{b}})}\BibitemShut
  {NoStop}%
\bibitem [{\citenamefont {Noguchi}(2022{\natexlab{a}})}]{nogu22a}%
  \BibitemOpen
  \bibfield  {author} {\bibinfo {author} {\bibfnamefont {H.}~\bibnamefont
  {Noguchi}},\ }\href {\doibase 10.1142/S021797922230002X} {\bibfield
  {journal} {\bibinfo  {journal} {Int. J. Mod. Phys. B}\ }\textbf {\bibinfo
  {volume} {36}},\ \bibinfo {pages} {2230002} (\bibinfo {year}
  {2022}{\natexlab{a}})}\BibitemShut {NoStop}%
\bibitem [{\citenamefont {Noguchi}(2016)}]{nogu16}%
  \BibitemOpen
  \bibfield  {author} {\bibinfo {author} {\bibfnamefont {H.}~\bibnamefont
  {Noguchi}},\ }\href {\doibase 10.1038/srep20935} {\bibfield  {journal}
  {\bibinfo  {journal} {Sci.\ Rep.}\ }\textbf {\bibinfo {volume} {6}},\
  \bibinfo {pages} {20935} (\bibinfo {year} {2016})}\BibitemShut {NoStop}%
\bibitem [{\citenamefont {Noguchi}(2022{\natexlab{b}})}]{nogu22b}%
  \BibitemOpen
  \bibfield  {author} {\bibinfo {author} {\bibfnamefont {H.}~\bibnamefont
  {Noguchi}},\ }\href {\doibase 10.1063/5.0098249} {\bibfield  {journal}
  {\bibinfo  {journal} {J. Chem. Phys.}\ }\textbf {\bibinfo {volume} {157}},\
  \bibinfo {pages} {034901} (\bibinfo {year} {2022}{\natexlab{b}})}\BibitemShut
  {NoStop}%
\bibitem [{\citenamefont {Noguchi}(2019)}]{nogu19c}%
  \BibitemOpen
  \bibfield  {author} {\bibinfo {author} {\bibfnamefont {H.}~\bibnamefont
  {Noguchi}},\ }\href {\doibase 10.1039/c9sm01622h} {\bibfield  {journal}
  {\bibinfo  {journal} {Soft Matter}\ }\textbf {\bibinfo {volume} {15}},\
  \bibinfo {pages} {8741} (\bibinfo {year} {2019})}\BibitemShut {NoStop}%
\bibitem [{\citenamefont {Noguchi}(2013)}]{nogu13}%
  \BibitemOpen
  \bibfield  {author} {\bibinfo {author} {\bibfnamefont {H.}~\bibnamefont
  {Noguchi}},\ }\href {\doibase 10.1209/0295-5075/102/68001} {\bibfield
  {journal} {\bibinfo  {journal} {EPL}\ }\textbf {\bibinfo {volume} {102}},\
  \bibinfo {pages} {68001} (\bibinfo {year} {2013})}\BibitemShut {NoStop}%
\bibitem [{\citenamefont {Gershon}\ \emph {et~al.}(2015)\citenamefont
  {Gershon}, \citenamefont {Breuer}, \citenamefont {Cohen}, \citenamefont
  {Cohrs}, \citenamefont {Gershon}, \citenamefont {Gilden}, \citenamefont
  {Grose}, \citenamefont {Hambleton}, \citenamefont {Kennedy}, \citenamefont
  {Oxman}, \citenamefont {Seward},\ and\ \citenamefont {Yamanishi}}]{gers15}%
  \BibitemOpen
  \bibfield  {author} {\bibinfo {author} {\bibfnamefont {A.~A.}\ \bibnamefont
  {Gershon}}, \bibinfo {author} {\bibfnamefont {J.}~\bibnamefont {Breuer}},
  \bibinfo {author} {\bibfnamefont {J.~I.}\ \bibnamefont {Cohen}}, \bibinfo
  {author} {\bibfnamefont {R.~J.}\ \bibnamefont {Cohrs}}, \bibinfo {author}
  {\bibfnamefont {M.~D.}\ \bibnamefont {Gershon}}, \bibinfo {author}
  {\bibfnamefont {D.}~\bibnamefont {Gilden}}, \bibinfo {author} {\bibfnamefont
  {C.}~\bibnamefont {Grose}}, \bibinfo {author} {\bibfnamefont
  {S.}~\bibnamefont {Hambleton}}, \bibinfo {author} {\bibfnamefont {P.~G.~E.}\
  \bibnamefont {Kennedy}}, \bibinfo {author} {\bibfnamefont {M.~N.}\
  \bibnamefont {Oxman}}, \bibinfo {author} {\bibfnamefont {J.~F.}\ \bibnamefont
  {Seward}}, \ and\ \bibinfo {author} {\bibfnamefont {K.}~\bibnamefont
  {Yamanishi}},\ }\href {\doibase 10.1038/nrdp.2015.16} {\bibfield  {journal}
  {\bibinfo  {journal} {Nat. Rev. Dis. Primers}\ }\textbf {\bibinfo {volume}
  {1}},\ \bibinfo {pages} {15016} (\bibinfo {year} {2015})}\BibitemShut
  {NoStop}%
\bibitem [{\citenamefont {Hsu}\ \emph {et~al.}(2018)\citenamefont {Hsu},
  \citenamefont {Lin}, \citenamefont {Harn}, \citenamefont {Hughes},
  \citenamefont {Tang},\ and\ \citenamefont {Yang}}]{hsu18}%
  \BibitemOpen
  \bibfield  {author} {\bibinfo {author} {\bibfnamefont {C.-K.}\ \bibnamefont
  {Hsu}}, \bibinfo {author} {\bibfnamefont {H.-H.}\ \bibnamefont {Lin}},
  \bibinfo {author} {\bibfnamefont {H.~I.-C.}\ \bibnamefont {Harn}}, \bibinfo
  {author} {\bibfnamefont {M.~W.}\ \bibnamefont {Hughes}}, \bibinfo {author}
  {\bibfnamefont {M.-J.}\ \bibnamefont {Tang}}, \ and\ \bibinfo {author}
  {\bibfnamefont {C.-C.}\ \bibnamefont {Yang}},\ }\href {\doibase
  10.1016/j.jdermsci.2018.03.004} {\bibfield  {journal} {\bibinfo  {journal}
  {J. Dermatol. Sci.}\ }\textbf {\bibinfo {volume} {90}},\ \bibinfo {pages}
  {232} (\bibinfo {year} {2018})}\BibitemShut {NoStop}%
\bibitem [{\citenamefont {Espa{\~n}ol}(1997)}]{espa97a}%
  \BibitemOpen
  \bibfield  {author} {\bibinfo {author} {\bibfnamefont {P.}~\bibnamefont
  {Espa{\~n}ol}},\ }\href@noop {} {\bibfield  {journal} {\bibinfo  {journal}
  {Europhys.\ Lett.}\ }\textbf {\bibinfo {volume} {40}},\ \bibinfo {pages}
  {631} (\bibinfo {year} {1997})}\BibitemShut {NoStop}%
\bibitem [{\citenamefont {Groot}\ and\ \citenamefont {Warren}(1997)}]{groo97}%
  \BibitemOpen
  \bibfield  {author} {\bibinfo {author} {\bibfnamefont {R.~D.}\ \bibnamefont
  {Groot}}\ and\ \bibinfo {author} {\bibfnamefont {P.~B.}\ \bibnamefont
  {Warren}},\ }\href@noop {} {\bibfield  {journal} {\bibinfo  {journal} {J.\
  Chem.\ Phys.}\ }\textbf {\bibinfo {volume} {107}},\ \bibinfo {pages} {4423}
  (\bibinfo {year} {1997})}\BibitemShut {NoStop}%
\bibitem [{\citenamefont {Shardlow}(2003)}]{shar03}%
  \BibitemOpen
  \bibfield  {author} {\bibinfo {author} {\bibfnamefont {T.}~\bibnamefont
  {Shardlow}},\ }\href@noop {} {\bibfield  {journal} {\bibinfo  {journal}
  {SIAM\ J.\ Sci.\ Comput.}\ }\textbf {\bibinfo {volume} {24}},\ \bibinfo
  {pages} {1267} (\bibinfo {year} {2003})}\BibitemShut {NoStop}%
\bibitem [{\citenamefont {Noguchi}\ and\ \citenamefont
  {Gompper}(2007)}]{nogu07a}%
  \BibitemOpen
  \bibfield  {author} {\bibinfo {author} {\bibfnamefont {H.}~\bibnamefont
  {Noguchi}}\ and\ \bibinfo {author} {\bibfnamefont {G.}~\bibnamefont
  {Gompper}},\ }\href {\doibase 10.1209/0295-5075/79/36002} {\bibfield
  {journal} {\bibinfo  {journal} {Europhys.\ Lett.}\ }\textbf {\bibinfo
  {volume} {78}},\ \bibinfo {pages} {36002} (\bibinfo {year}
  {2007})}\BibitemShut {NoStop}%
\bibitem [{\citenamefont {Hannezo}\ \emph {et~al.}(2014)\citenamefont
  {Hannezo}, \citenamefont {Prost},\ and\ \citenamefont {Joanny}}]{hann14}%
  \BibitemOpen
  \bibfield  {author} {\bibinfo {author} {\bibfnamefont {E.}~\bibnamefont
  {Hannezo}}, \bibinfo {author} {\bibfnamefont {J.}~\bibnamefont {Prost}}, \
  and\ \bibinfo {author} {\bibfnamefont {J.-F.}\ \bibnamefont {Joanny}},\
  }\href {\doibase 10.1073/pnas.1312076111} {\bibfield  {journal} {\bibinfo
  {journal} {Proc.\ Natl.\ Acad.\ Sci.\ USA}\ }\textbf {\bibinfo {volume}
  {111}},\ \bibinfo {pages} {27} (\bibinfo {year} {2014})}\BibitemShut
  {NoStop}%
\bibitem [{\citenamefont {Marmottant}\ \emph {et~al.}(2009)\citenamefont
  {Marmottant}, \citenamefont {Mgharbel}, \citenamefont {K\"afer},
  \citenamefont {Audren}, \citenamefont {Rieu}, \citenamefont {Vial},
  \citenamefont {van~der Sanden}, \citenamefont {Mar\'ee}, \citenamefont
  {Graner},\ and\ \citenamefont {Delano\"e-Ayari}}]{marm09}%
  \BibitemOpen
  \bibfield  {author} {\bibinfo {author} {\bibfnamefont {P.}~\bibnamefont
  {Marmottant}}, \bibinfo {author} {\bibfnamefont {A.}~\bibnamefont
  {Mgharbel}}, \bibinfo {author} {\bibfnamefont {J.}~\bibnamefont {K\"afer}},
  \bibinfo {author} {\bibfnamefont {B.}~\bibnamefont {Audren}}, \bibinfo
  {author} {\bibfnamefont {J.-P.}\ \bibnamefont {Rieu}}, \bibinfo {author}
  {\bibfnamefont {J.-C.}\ \bibnamefont {Vial}}, \bibinfo {author}
  {\bibfnamefont {B.}~\bibnamefont {van~der Sanden}}, \bibinfo {author}
  {\bibfnamefont {A.~F.~M.}\ \bibnamefont {Mar\'ee}}, \bibinfo {author}
  {\bibfnamefont {F.}~\bibnamefont {Graner}}, \ and\ \bibinfo {author}
  {\bibfnamefont {H.}~\bibnamefont {Delano\"e-Ayari}},\ }\href {\doibase
  10.1073/pnas.0902085106} {\bibfield  {journal} {\bibinfo  {journal} {Proc.\
  Natl.\ Acad.\ Sci.\ USA}\ }\textbf {\bibinfo {volume} {106}},\ \bibinfo
  {pages} {17271} (\bibinfo {year} {2009})}\BibitemShut {NoStop}%
\bibitem [{\citenamefont {Truesdell}(1983)}]{true83}%
  \BibitemOpen
  \bibfield  {author} {\bibinfo {author} {\bibfnamefont {C.}~\bibnamefont
  {Truesdell}},\ }\href
  {https://www.ams.org/journals/bull/1983-09-03/S0273-0979-1983-15187-X/}
  {\bibfield  {journal} {\bibinfo  {journal} {Bull. Am. Math. Soc.}\ }\textbf
  {\bibinfo {volume} {9}},\ \bibinfo {pages} {293} (\bibinfo {year}
  {1983})}\BibitemShut {NoStop}%
\bibitem [{\citenamefont {Noguchi}(2011)}]{nogu11a}%
  \BibitemOpen
  \bibfield  {author} {\bibinfo {author} {\bibfnamefont {H.}~\bibnamefont
  {Noguchi}},\ }\href {\doibase 10.1103/PhysRevE.83.061919} {\bibfield
  {journal} {\bibinfo  {journal} {Phys. Rev. E}\ }\textbf {\bibinfo {volume}
  {83}},\ \bibinfo {pages} {061919} (\bibinfo {year} {2011})}\BibitemShut
  {NoStop}%
\bibitem [{\citenamefont {Nakagawa}\ and\ \citenamefont
  {Noguchi}(2018)}]{naka18}%
  \BibitemOpen
  \bibfield  {author} {\bibinfo {author} {\bibfnamefont {K.~M.}\ \bibnamefont
  {Nakagawa}}\ and\ \bibinfo {author} {\bibfnamefont {H.}~\bibnamefont
  {Noguchi}},\ }\href {\doibase 10.1039/c7sm02326j} {\bibfield  {journal}
  {\bibinfo  {journal} {Soft Matter}\ }\textbf {\bibinfo {volume} {14}},\
  \bibinfo {pages} {1397} (\bibinfo {year} {2018})}\BibitemShut {NoStop}%
\bibitem [{\citenamefont {Frisch}\ and\ \citenamefont
  {Screaton}(2001)}]{fris01}%
  \BibitemOpen
  \bibfield  {author} {\bibinfo {author} {\bibfnamefont {S.~M.}\ \bibnamefont
  {Frisch}}\ and\ \bibinfo {author} {\bibfnamefont {R.~A.}\ \bibnamefont
  {Screaton}},\ }\href {\doibase 10.1016/S0955-0674(00)00251-9} {\bibfield
  {journal} {\bibinfo  {journal} {Curr. Opin. Cell Biol.}\ }\textbf {\bibinfo
  {volume} {13}},\ \bibinfo {pages} {555} (\bibinfo {year} {2001})}\BibitemShut
  {NoStop}%
\bibitem [{\citenamefont {Grossmann}(2002)}]{gros02}%
  \BibitemOpen
  \bibfield  {author} {\bibinfo {author} {\bibfnamefont {J.}~\bibnamefont
  {Grossmann}},\ }\href {\doibase 10.1023/A:1015312119693} {\bibfield
  {journal} {\bibinfo  {journal} {Apoptosis}\ }\textbf {\bibinfo {volume}
  {7}},\ \bibinfo {pages} {247} (\bibinfo {year} {2002})}\BibitemShut {NoStop}%
\bibitem [{\citenamefont {de~Gennes}\ \emph {et~al.}(2003)\citenamefont
  {de~Gennes}, \citenamefont {Brochard-Wyart},\ and\ \citenamefont
  {Quere}}]{dege03}%
  \BibitemOpen
  \bibfield  {author} {\bibinfo {author} {\bibfnamefont {P.~G.}\ \bibnamefont
  {de~Gennes}}, \bibinfo {author} {\bibfnamefont {F.}~\bibnamefont
  {Brochard-Wyart}}, \ and\ \bibinfo {author} {\bibfnamefont {D.}~\bibnamefont
  {Quere}},\ }\href@noop {} {\emph {\bibinfo {title} {Capillarity and wetting
  phenomena: Drops, bubbles, perls, waves}}}\ (\bibinfo  {publisher}
  {Springer},\ \bibinfo {address} {New York},\ \bibinfo {year}
  {2003})\BibitemShut {NoStop}%
\bibitem [{\citenamefont {Utada}\ \emph {et~al.}(2005)\citenamefont {Utada},
  \citenamefont {Lorenceau}, \citenamefont {Link}, \citenamefont {Kaplan},
  \citenamefont {Stone},\ and\ \citenamefont {Weitz}}]{utad05}%
  \BibitemOpen
  \bibfield  {author} {\bibinfo {author} {\bibfnamefont {A.~S.}\ \bibnamefont
  {Utada}}, \bibinfo {author} {\bibfnamefont {E.}~\bibnamefont {Lorenceau}},
  \bibinfo {author} {\bibfnamefont {D.~R.}\ \bibnamefont {Link}}, \bibinfo
  {author} {\bibfnamefont {P.~D.}\ \bibnamefont {Kaplan}}, \bibinfo {author}
  {\bibfnamefont {H.~A.}\ \bibnamefont {Stone}}, \ and\ \bibinfo {author}
  {\bibfnamefont {D.~A.}\ \bibnamefont {Weitz}},\ }\href {\doibase
  10.1126/science.1109164} {\bibfield  {journal} {\bibinfo  {journal}
  {Science}\ }\textbf {\bibinfo {volume} {308}},\ \bibinfo {pages} {537}
  (\bibinfo {year} {2005})}\BibitemShut {NoStop}%
\bibitem [{\citenamefont {McEvoy}\ \emph {et~al.}(2022)\citenamefont {McEvoy},
  \citenamefont {Sneh}, \citenamefont {Moeendarbary}, \citenamefont
  {Javanmardi}, \citenamefont {Efimova}, \citenamefont {Yang}, \citenamefont
  {Marino-Bravante}, \citenamefont {Chen}, \citenamefont {Escribano},
  \citenamefont {Spill}, \citenamefont {Garcia-Aznar}, \citenamefont
  {Weeraratna}, \citenamefont {Svitkina}, \citenamefont {Kamm},\ and\
  \citenamefont {Shenoy}}]{mcev22}%
  \BibitemOpen
  \bibfield  {author} {\bibinfo {author} {\bibfnamefont {E.}~\bibnamefont
  {McEvoy}}, \bibinfo {author} {\bibfnamefont {T.}~\bibnamefont {Sneh}},
  \bibinfo {author} {\bibfnamefont {E.}~\bibnamefont {Moeendarbary}}, \bibinfo
  {author} {\bibfnamefont {Y.}~\bibnamefont {Javanmardi}}, \bibinfo {author}
  {\bibfnamefont {N.}~\bibnamefont {Efimova}}, \bibinfo {author} {\bibfnamefont
  {C.}~\bibnamefont {Yang}}, \bibinfo {author} {\bibfnamefont {G.~E.}\
  \bibnamefont {Marino-Bravante}}, \bibinfo {author} {\bibfnamefont
  {X.}~\bibnamefont {Chen}}, \bibinfo {author} {\bibfnamefont {J.}~\bibnamefont
  {Escribano}}, \bibinfo {author} {\bibfnamefont {F.}~\bibnamefont {Spill}},
  \bibinfo {author} {\bibfnamefont {J.~M.}\ \bibnamefont {Garcia-Aznar}},
  \bibinfo {author} {\bibfnamefont {A.~T.}\ \bibnamefont {Weeraratna}},
  \bibinfo {author} {\bibfnamefont {T.~M.}\ \bibnamefont {Svitkina}}, \bibinfo
  {author} {\bibfnamefont {R.~D.}\ \bibnamefont {Kamm}}, \ and\ \bibinfo
  {author} {\bibfnamefont {V.~B.}\ \bibnamefont {Shenoy}},\ }\href {\doibase
  10.1038/s41467-022-34701-y} {\bibfield  {journal} {\bibinfo  {journal} {Nat.
  Commun.}\ }\textbf {\bibinfo {volume} {13}},\ \bibinfo {pages} {7089}
  (\bibinfo {year} {2022})}\BibitemShut {NoStop}%
\bibitem [{\citenamefont {Voorhees}(1985)}]{voor85}%
  \BibitemOpen
  \bibfield  {author} {\bibinfo {author} {\bibfnamefont {P.~W.}\ \bibnamefont
  {Voorhees}},\ }\href {\doibase 10.1007/BF01017860} {\bibfield  {journal}
  {\bibinfo  {journal} {J. Stat. Phys.}\ }\textbf {\bibinfo {volume} {38}},\
  \bibinfo {pages} {231} (\bibinfo {year} {1985})}\BibitemShut {NoStop}%
\bibitem [{\citenamefont {Kabalnov}(2001)}]{kaba01}%
  \BibitemOpen
  \bibfield  {author} {\bibinfo {author} {\bibfnamefont {A.}~\bibnamefont
  {Kabalnov}},\ }\href {\doibase 10.1081/DIS-100102675} {\bibfield  {journal}
  {\bibinfo  {journal} {J. Disper. Sci. Technol.}\ }\textbf {\bibinfo {volume}
  {22}},\ \bibinfo {pages} {1} (\bibinfo {year} {2001})}\BibitemShut {NoStop}%
\bibitem [{\citenamefont {Watanabe}\ \emph {et~al.}(2014)\citenamefont
  {Watanabe}, \citenamefont {Suzuki}, \citenamefont {Inaoka},\ and\
  \citenamefont {Ito}}]{wata14}%
  \BibitemOpen
  \bibfield  {author} {\bibinfo {author} {\bibfnamefont {H.}~\bibnamefont
  {Watanabe}}, \bibinfo {author} {\bibfnamefont {M.}~\bibnamefont {Suzuki}},
  \bibinfo {author} {\bibfnamefont {H.}~\bibnamefont {Inaoka}}, \ and\ \bibinfo
  {author} {\bibfnamefont {N.}~\bibnamefont {Ito}},\ }\href {\doibase
  10.1063/1.4903811} {\bibfield  {journal} {\bibinfo  {journal} {J. Chem.
  Phys.}\ }\textbf {\bibinfo {volume} {141}},\ \bibinfo {pages} {234703}
  (\bibinfo {year} {2014})}\BibitemShut {NoStop}%
\bibitem [{\citenamefont {Schweikart}\ and\ \citenamefont
  {Fery}(2009)}]{schw09}%
  \BibitemOpen
  \bibfield  {author} {\bibinfo {author} {\bibfnamefont {A.}~\bibnamefont
  {Schweikart}}\ and\ \bibinfo {author} {\bibfnamefont {A.}~\bibnamefont
  {Fery}},\ }\href {\doibase 10.1007/s00604-009-0153-3} {\bibfield  {journal}
  {\bibinfo  {journal} {Microchim Acta}\ }\textbf {\bibinfo {volume} {165}},\
  \bibinfo {pages} {249} (\bibinfo {year} {2009})}\BibitemShut {NoStop}%
\bibitem [{\citenamefont {Li}\ \emph {et~al.}(2012)\citenamefont {Li},
  \citenamefont {Cao}, \citenamefont {Feng},\ and\ \citenamefont {Gao}}]{li12}%
  \BibitemOpen
  \bibfield  {author} {\bibinfo {author} {\bibfnamefont {B.}~\bibnamefont
  {Li}}, \bibinfo {author} {\bibfnamefont {Y.-P.}\ \bibnamefont {Cao}},
  \bibinfo {author} {\bibfnamefont {X.-Q.}\ \bibnamefont {Feng}}, \ and\
  \bibinfo {author} {\bibfnamefont {H.}~\bibnamefont {Gao}},\ }\href {\doibase
  10.1039/c2sm00011c} {\bibfield  {journal} {\bibinfo  {journal} {Soft Matter}\
  }\textbf {\bibinfo {volume} {8}},\ \bibinfo {pages} {5728} (\bibinfo {year}
  {2012})}\BibitemShut {NoStop}%
\bibitem [{\citenamefont {Deserno}(2009)}]{dese09}%
  \BibitemOpen
  \bibfield  {author} {\bibinfo {author} {\bibfnamefont {M.}~\bibnamefont
  {Deserno}},\ }\href {\doibase 10.1002/marc.200900090} {\bibfield  {journal}
  {\bibinfo  {journal} {Macromol.\ Rapid.\ Commun.}\ }\textbf {\bibinfo
  {volume} {30}},\ \bibinfo {pages} {752} (\bibinfo {year} {2009})}\BibitemShut
  {NoStop}%
\bibitem [{\citenamefont {Shiba}\ and\ \citenamefont {Noguchi}(2011)}]{shib11}%
  \BibitemOpen
  \bibfield  {author} {\bibinfo {author} {\bibfnamefont {H.}~\bibnamefont
  {Shiba}}\ and\ \bibinfo {author} {\bibfnamefont {H.}~\bibnamefont
  {Noguchi}},\ }\href {\doibase 10.1103/PhysRevE.84.031926} {\bibfield
  {journal} {\bibinfo  {journal} {Phys. Rev. E}\ }\textbf {\bibinfo {volume}
  {84}},\ \bibinfo {pages} {031926} (\bibinfo {year} {2011})}\BibitemShut
  {NoStop}%
\bibitem [{\citenamefont {Hu}\ \emph {et~al.}(2013)\citenamefont {Hu},
  \citenamefont {Diggins},\ and\ \citenamefont {Deserno}}]{hu13a}%
  \BibitemOpen
  \bibfield  {author} {\bibinfo {author} {\bibfnamefont {M.}~\bibnamefont
  {Hu}}, \bibinfo {author} {\bibfnamefont {P.}~\bibnamefont {Diggins}}, \ and\
  \bibinfo {author} {\bibfnamefont {M.}~\bibnamefont {Deserno}},\ }\href
  {\doibase 10.1063/1.4808077} {\bibfield  {journal} {\bibinfo  {journal} {J.
  Chem. Phys.}\ }\textbf {\bibinfo {volume} {138}},\ \bibinfo {pages} {214110}
  (\bibinfo {year} {2013})}\BibitemShut {NoStop}%
\bibitem [{\citenamefont {Tolpekina}\ \emph {et~al.}(2004)\citenamefont
  {Tolpekina}, \citenamefont {{den Otter}},\ and\ \citenamefont
  {Briels}}]{tolp04}%
  \BibitemOpen
  \bibfield  {author} {\bibinfo {author} {\bibfnamefont {T.~V.}\ \bibnamefont
  {Tolpekina}}, \bibinfo {author} {\bibfnamefont {W.~K.}\ \bibnamefont {{den
  Otter}}}, \ and\ \bibinfo {author} {\bibfnamefont {W.~J.}\ \bibnamefont
  {Briels}},\ }\href {\doibase 10.1063/1.1796254} {\bibfield  {journal}
  {\bibinfo  {journal} {J.\ Chem.\ Phys.}\ }\textbf {\bibinfo {volume} {121}},\
  \bibinfo {pages} {8014} (\bibinfo {year} {2004})}\BibitemShut {NoStop}%
\end{thebibliography}

%merlin.mbs apsrev4-1.bst 2010-07-25 4.21a (PWD, AO, DPC) hacked
%Control: key (0)
%Control: author (72) initials jnrlst
%Control: editor formatted (1) identically to author
%Control: production of article title (-1) disabled
%Control: page (0) single
%Control: year (1) truncated
%Control: production of eprint (0) enabled
%

\end{document}